\documentclass[11pt]{JHEP3}
\usepackage{graphicx}
\usepackage{epsfig}
\usepackage{amssymb}
\usepackage{amsmath}
\usepackage{comment} 
\newcommand{\bea}{\begin{eqnarray}}
\newcommand{\eea}{\end{eqnarray}}
\newcommand{\beq}{\begin{equation}}
\newcommand{\eeq}{\end{equation}}
\newcommand{\nn}{\nonumber}

\newcommand{\C}[1]{{\mathcal{#1}}}
\newcommand{\B}[1]{{\mathbf{#1}}}
\newcommand{\R}[1]{{\mathrm{#1}}}
\newcommand{\BB}[1]{{\mathbb{#1}}}

\newcommand{\half}{{\frac{1}{2}}}

\newcommand{\quarter}{{\frac{1}{4}}}
\newcommand{\third}{{\frac{1}{3}}}

\newcommand{\ket}[1]{\,\vert #1\rangle}
\newcommand{\bra}[1]{\langle #1\vert\,}
\newcommand{\braket}[2]{\langle #1\vert\,#2\rangle}

\newcommand{\avg}[1]{\left\langle #1 \right\rangle}

\newcommand{\Z}{{\C Z}}
\newcommand{\Tr}{\mathrm{Tr}}
\newcommand{\res}{\mathrm{Res}}

\newcommand{\tW}{{\widetilde W}}

\newcommand{\Oh}{{\C O}}

\newenvironment{property}[2][Property]{\begin{trivlist}
\item[\hskip \labelsep {\bfseries #1 #2 (P#2):}]}{\end{trivlist}}

\newcommand{\qed}{\nobreak \ifvmode \relax \else
      \ifdim\lastskip<1.5em \hskip-\lastskip
      \hskip1.5em plus0em minus0.5em \fi \nobreak
      \vrule height0.75em width0.5em depth0.25em\fi}

\title{The Spectrum of FZZT Branes Beyond the Planar Limit}
\author{Max R. Atkin\\Rudolf Peierls Center for Theoretical Physics\\Department of Physics, 1 Keble Road, Oxford OX1 3NP, UK\\ E-mail: \email{ m.atkin1@physics.ox.ac.uk}}
\author{John F. Wheater\\Rudolf Peierls Center for Theoretical Physics\\Department of Physics, 1 Keble Road, Oxford OX1 3NP, UK\\ E-mail: \email{ j.wheater@physics.ox.ac.uk}}

\abstract{Minimal string theory has a number of FZZT brane boundary states; one for each Cardy state of the minimal model. It was conjectured by Seiberg and Shih that all branes in a minimal string theory could be expressed as a linear combination of the brane associated to the identity operator of the minimal model with complex shifts in the boundary cosmological constant. Subsequently it was found that this identification of FZZT branes does not hold exactly for some cylinder amplitudes but was spoiled by terms that are associated with vanishing worldsheet area and are therefore non-universal. In this paper we investigate this claim systematically, using both Liouville and matrix model methods, beyond the planar limit. We find that the aforementioned identification of FZZT branes is spoiled by terms that do not admit an interpretation as non-universal terms. Furthermore, the spoiling terms as computed using the matrix model are found to be in agreement with those coming from Liouville theory, which also suggests that these terms have universal meaning. Finally, we also investigate the identification of FZZT branes by replacing the boundary state with a sum of local operators. We find in this case that the brane associated with the identity operator appears to be special as it is the only one to correctly reproduce the correlation numbers for bulk operators on the torus.}
\preprint{}

\begin{document}
\section{Introduction}
Minimal string theories are obtained by interpreting Liouville theory coupled to a $(p,q)$ minimal model as the worldsheet theory for a string. Advances in the last decade \cite{Fateev:2000ik}\cite{Zamolodchikov:2001ah}\cite{Ponsot:2001ng} have identified the consistent boundary states of the Liouville theory and hence, in the string interpretation, allowed the branes to be constructed. The boundary conditions in Liouville theory can be classified into two families, FZZT and ZZ. The FZZT boundary states $\ket{\sigma}_{\R{FZZT}}$ are labelled by a continuous parameter, $\sigma$, which is related to the boundary cosmological constant. The ZZ boundary conditions, which we will not be considering in this paper, are labelled by a pair of integers $(m,n)$. To obtain a brane in minimal string theory we must also specify a boundary condition for the matter theory. The consistent boundary states $\ket{k,l}$ for the minimal models are in one-to-one correspondence with the primary fields and so are labelled by the Kac indices $(k,l)$. To obtain a brane in the minimal string we simply take a tensor product of these two states and so, for example, the FZZT branes are given by $\ket{\sigma;k,l}=\ket{\sigma}_{\R{FZZT}}\otimes \ket{k,l}$.

Once amplitudes for a disk with FZZT boundary conditions and one vertex operator in the bulk had been computed it was noticed in \cite{Seiberg:2003nm} that the construction just outlined appears to over count the number of FZZT branes.  Up to null states it seems that 
\beq \label{SS1}
\ket{\sigma;k,l}=\sum^{k-1}_{n=-(k-1),2} \sum^{l-1}_{m=-(l-1),2}  \ket{\sigma+i\frac{mp+nq}{\sqrt{pq}};1,1}.
\eeq
In effect the disc amplitudes behave as though the information about the matter sector boundary condition can be subsumed in complex shifts of the boundary cosmological constant which describes the Liouville sector boundary condition. It was noted in \cite{Kutasov:2004fg} that the deviation from the Seiberg-Shih relations for some cylinder amplitudes is non-zero but it was argued that the deviation should be discarded as it only appears for certain disk areas or boundary lengths.

In this paper we test the conjecture \eqref{SS1} systematically, first using general cylinder amplitudes where we show that the prescription described above for discarding deviations is inadequate; this case is discussed in greater detail in a recent paper \cite{Gesser:2010fi}. We then conjecture a new prescription which is correct at cylinder level and test it by considering amplitudes for disc with one handle. Disk amplitudes at higher genus have not been calculated in Liouville theory so we turn instead to the matrix model formulation. In general this approach suffers from the difficulty of constructing the full family of FZZT branes which arise from the FZZT boundary state tensored with \emph{any} matter boundary state, although there has been recent progress in this direction \cite{Ishiki:2010wb, Bourgine:2010ja}. There is however a special case where the construction of all FZZT branes is easy; the $(3,4)$ minimal string which corresponds to the critical point of the Ising model coupled to Liouville gravity.  We use this to show that the conjecture is in fact false.  We also show that the failure of our conjecture is equivalent to being unable to replace branes other than  the fixed spin boundary with a sum of local operators. As a by-product of our investigations  we show there is an easy way to extract correlation numbers from loop amplitudes and verify its correctness by reproducing the results of \cite{Belavin:2010pj} for the $(2,5)$ minimal string.

This paper is arranged as follows. In Section \ref{msandbranes} we review the $(p,q)$ minimal string, FZZT brane, and the motivation for the identification \eqref{SS1} in addition to introducing our conjecture. In Section \ref{CylinderSec} we examine the failure of the Seiberg-Shih relations for $(p,q)$ cylinder amplitudes with arbitrary FZZT brane boundary states and then motivate our conjecture.  In Section \ref{MatrixSec} we review the calculation of matrix model amplitudes and the results we need for cylinder and disc-with-handle amplitudes. We then examine the Seiberg-Shih deviation for the disc-with-handle amplitudes and test our conjecture. In Section \ref{Localstates} we investigate the deviations by expressing boundary states as sums of local operators and then studying how the Seiberg-Shih relations act on this expansion.  Finally, given that our results indicate that the different branes are not equivalent, we discuss in the conclusion whether we can identify in the matrix model all the dual branes. Various technical results are given in the appendices.


\section{$(p,q)$ Minimal Strings and FZZT branes\label{msandbranes}}
The Liouville theory on a manifold with boundary is defined by the Lagrangian,
\bea S_L=\frac{1}{4\pi}\int_M d^2x\sqrt{g}\left(g^{ab}\partial_a\phi\partial_b\phi+Q\phi R[g]+4\pi\mu e^{2b\phi}\right) + \int_{\partial M} \mu_B e^{b\phi} dx \eea
where $Q=b^{-1}+b$, $\mu$ is the bulk cosmological constant and $\mu_B$ is the boundary cosmological constant. The primary fields of the theory are $V_{\alpha}=e^{2\alpha\phi}$. The worldsheet theory for the minimal string is obtained by coupling Liouville theory to a $(p,q)$ minimal model; this requires that $b^2=\frac{p}{q}$. 

The physical operators of the $(p,q)$ minimal string were first given by \cite{Lian:1992aj} as $\C{U}_n \equiv \C{O}_n V_{\alpha(n)}$ where,
\bea
\alpha(n) =b \frac{p+q-n}{2p} \quad n \neq 0 \mod p \quad n \neq 0 \mod q \quad n \in \BB{Z}
\eea
and $\C{O}_n$ is an operator containing contributions from the ghost and matter sector in addition to derivatives of the Liouville field. An important physical operator is obtained by gravitationally dressing the primary fields $\Oh_{r,s}$ of the minimal model; such operators are referred to as tachyons and take the form,
\bea \C{T}_{r,s}=&&c\bar{c}\Oh_{r,s}V_{\alpha_{r,s}} \\ 
\alpha_{r,s} =&& \frac{p+q-|qr-ps|}{2\sqrt{pq}}\eea
 where $c$ and $\bar{c}$ are ghost fields. The full conformal bootstrap of Liouville theory on the disc was performed in \cite{Fateev:2000ik}\cite{Ponsot:2001ng}, resulting in the identification of a consistent boundary condition for the Liouville theory; the FZZT brane. As part of the conformal bootstrap, the bulk one-point function on a disc, i.e. the bulk one-point function with FZZT boundary, was calculated:
\beq
\label{FZZTstate} \Psi_{\sigma_1}(P) \equiv \langle v_P |\sigma\rangle_{FZZT}=(\pi \mu \gamma(b^2))^ {-iP/b}\Gamma(1+2ibP)\Gamma(1+2iP/b)\frac{\cos(2\pi \sigma P)}{iP}
\eeq 
where $\sigma$ is defined implicitly by, $\mu \cosh^2\pi b \sigma = {\mu_B}^2\sin(\pi b^2)$ and $\ket{v_P}$ is the primary state associated to $V_{Q/2+iP}$.
In the minimal string the full FZZT brane is $\ket{\sigma;k,l}=|\sigma\rangle_{FZZT} \otimes |k,l\rangle$, where $|k,l\rangle$ is the $(k,l)$ Cardy boundary state of the minimal model. We will find it convenient to follow \cite{Seiberg:2003nm} in rescaling the bulk and boundary cosmological constant so that $\pi \mu \gamma(b^2) \rightarrow \mu$ and $\mu_B \sqrt{\pi\gamma(b^2)\sin(\pi b^2)} \rightarrow \mu_B$. Furthermore, it was noted in \cite{Dorn:1994xn} \cite{Zamolodchikov:1995aa} that the results of the conformal bootstrap on the sphere are invariant under the duality transformation $b\rightarrow \frac{1}{b}$, $\mu \rightarrow \tilde{\mu}=\mu^{\frac{1}{b^2}}$. This self duality is present in the bootstrap on the disc if the boundary cosmological constant transforms as, $\mu_B \rightarrow \tilde{\mu}_B$, where $\frac{\tilde{\mu}_B}{\sqrt{\tilde{\mu}}}=\cosh \frac{\pi \sigma}{b}$. The transformed boundary condition is referred to as the dual FZZT brane and provides a physically equivalent description of the FZZT brane.

The amplitude in the minimal string for the insertion of a tachyon in the bulk of a disc is obtained using \eqref{FZZTstate} \cite{Seiberg:2003nm},
\bea\label{tachyonDisk}
\braket{\C{T}_{r,s}}{\sigma;k,l}=&&A_D\left[\sin\left(\frac{\pi t}{p}\right)\sin\left(\frac{\pi t}{q}\right)\right]^{\frac{1}{2}} \Gamma(2b \alpha -b^2) \Gamma(2b^{-1} \alpha -b^{-2}-1) \times\\
&&U_{k-1}\left(\cos\frac{\pi t}{p}\right)U_{l-1}\left(\cos\frac{\pi t}{q}\right)\mu^\frac{t}{2p}\cosh\left(\frac{\pi t \sigma}{\sqrt{pq}}\right) \nn
\eea 
where $t=qr-ps$, $2\alpha=Q-\frac{t}{\sqrt{pq}}$, $A_D$ is a constant independent of $k$, $r$, $s$ and $l$ and $U_n(x)$ is the $n$th Chebyshev polynomial of the second kind. This expressions for the disc amplitudes motivated Seiberg and Shih to conjecture in \cite{Seiberg:2003nm} that the construction just outlined overcounts the number of FZZT branes and that, up to BRST null states, the following identification holds,
\beq \label{SSequiv}
\ket{\sigma;k,l}=\sum^{k-1}_{n=-(k-1),2} \sum^{l-1}_{m=-(l-1),2}  \ket{\sigma+i\frac{mp+nq}{\sqrt{pq}};1,1}.
\eeq 
This relation, which we will refer to as the SS (for Seiberg-Shih) relation, was originally obtained by inspection of \eqref{tachyonDisk} but  was later derived for discs using the ground ring in \cite{Basu:2005sda}. Essentially it states that there is only a single FZZT brane $\ket{\sigma;1,1}$ and that all the others are related to it by complex shifts in the boundary cosmological constant which has somehow absorbed all the information about the matter boundary condition.

The validity of the SS relation has been checked for the discs and cylinder amplitudes \cite{Kutasov:2004fg}\cite{Ambjorn}\cite{Gesser:2010fi} and is known to fail in some cases of the latter. For our purposes it is useful to define the deviation from the SS relations for a particular amplitude $\C{A}(k,l;\sigma|X)$ as
\beq 
\Delta \C{A}(k,l;\sigma|X)=\C{A}(k,l;\sigma|X)-\sum^{k-1}_{n=-(k-1),2} \sum^{l-1}_{m=-(l-1),2}  \C{A}(1,1;\sigma+i\frac{mp+nq}{\sqrt{pq}}|X)
\eeq
where the presence of other boundaries and operator insertions is denoted by $X$. It was first argued in \cite{Kutasov:2004fg} that the deviations noticed there should be discarded as they have the following properties:

\begin{property}{1}
The dependence of the deviation on the bulk cosmological constant could be cancelled by setting the IR cutoff (that was needed in order to compute the amplitude) to the volume of the Liouville direction.  \label{prop1}
\end{property}

\begin{property}{2}
The inverse Laplace transform, with respect to all boundary cosmological constants, of the deviation minus any regularisation dependent parts is zero almost everywhere i.e. it is supported only at points.   \label{prop2}
\end{property}

Essentially P1 says that it can be arranged for the deviation to occur only for surfaces of zero area while P2 says that the deviation is non-zero only for particular values of the boundary length; and it is argued that in neither case can the deviation have any physical meaning. 

In Section \ref{CylinderSec} we examine the SS relations for cylinder amplitudes with arbitrary FZZT brane boundary states. These amplitudes are given in \cite{Ambjorn} in a rather unwieldy form; we give them in Appendix \ref{Cylindercalc} in a more usable form similar to that given recently by \cite{Gesser:2010fi}. We will show that there exist choices of boundaries for which the deviation  possesses neither P1 nor P2. 
The SS relations must therefore be modified or discarded. This leads us to define the property:
\begin{property}{3}
\label{prop3}
The only worldsheet geometries that contribute to the deviation are those with particular values of boundary length (i.e. zero) {\emph{and}} their  duals. \end{property}

P3 means that all deviations may be written as the sum of two terms that are dual to one another such that the inverse Laplace transform of each term with respect to the boundary cosmological constant and dual boundary cosmological constant respectively, vanishes almost everywhere. We conjecture that all deviations possessing P3 should be discarded (or equivalently that the SS relations for amplitudes hold up to deviations having P3). In the rest of this paper we investigate this conjecture.


\section{FZZT-FZZT Cylinder Amplitudes}
\label{CylinderSec}
In this section we use the cylinder amplitude between arbitrary FZZT branes, which we denote $\C{Z}(k,l;\sigma_1|r,s;\sigma_2)$ to show that neither P1 nor P2  hold for all cylinder amplitudes. First consider a specific example, the case of the $(3,4)$ minimal string. The general amplitude in Appendix \ref{Cylindercalc} gives the following results \footnote{These expressions are correct up to an unimportant common normalisation constant and additions of numerical constants.},
\bea
\label{34amps}
&&\C{Z}(1,1;\sigma_1|1,1;\sigma_2)=\C{Z}(2,1;\sigma_1|2,1;\sigma_2)=\ln(\frac{z_1-z_2}{x_1-x_2})+\frac{1}{2\sqrt{3}\epsilon}, \\
&&\C{Z}(2,1;\sigma_1|1,1;\sigma_2)=-\ln(z_1+z_2)+\frac{1}{4\sqrt{3}\epsilon}, \nn\\
&&\C{Z}(1,1;\sigma_1|1,2;\sigma_2)=\C{Z}(2,1;\sigma_1|1,2;\sigma_2)=-\ln(-1 + 2 z_1^2 + 
  2 \sqrt{2} z_1 z_2 + 2 z_2^2)+\frac{1}{2\sqrt{3}\epsilon},\nn\\
&&\C{Z}(1,2;\sigma_1|1,2;\sigma_2)=\ln(\frac{z_1-z_2}{(z_1+z_2)(x_1-x_2)})+\frac{3}{4\sqrt{3}\epsilon},\nn
\eea
where $\epsilon$ is a regulator introduced to cut-off an IR divergence present in the integral and we have introduced the following notation which will be useful in the remainder of the paper,
\bea
z \equiv \cosh\left(\frac{\pi \sigma}{\sqrt{pq}}\right), \qquad x \equiv \frac{\mu_B}{\sqrt{\mu}} = \cosh\left(\frac{\pi p \sigma}{\sqrt{pq}}\right), \qquad \tilde{x} \equiv \frac{\tilde{\mu}_B}{\sqrt{\tilde{\mu}}} = \cosh\left(\frac{\pi q \sigma}{\sqrt{pq}}\right).
\eea
Clearly we need to remove the dependence on the regulator and in \cite{Kutasov:2004fg} two possibilities were considered; either throw away the terms that diverge as $\epsilon$ goes to zero (this gives agreement with the matrix model), or set it equal to the volume of the Liouville direction. If we choose to do the latter then P1 should apply, if we choose the former then P2 should apply.  As an example of a deviation which satisfies P1 and P2, consider,
\bea
\label{exampleDev}
\Delta\C{Z}(2,1;\sigma_1|1,1;\sigma_2) = -\frac{\sqrt{3}}{4\epsilon}-\log(x_1+x_2).
\eea
Choosing $\epsilon$ equal to the volume of the Liouville direction, $\frac{1}{\epsilon} = \frac{1}{b}\log\frac{\Lambda}{\mu}$ where $\Lambda$ is a constant and recalling that $x = \mu_B/\sqrt{\mu}$ we find
\bea
\Delta\C{Z}(2,1;\sigma_1|1,1;\sigma_2) = -\frac{1}{2}\log\Lambda-\log({\mu_B}_1+{\mu_B}_2).
\eea
Since this is independent of $\mu$ it was interpreted as non-universal and therefore set to zero. To see that \eqref{exampleDev} also satisfies property P2, we take the inverse Laplace transform to get 
\bea
\Delta\C{Z}(2,1;L_1|1,1;L_2) = -\frac{\sqrt{3}}{4\epsilon}\delta(L_1)\delta(L_2) +\delta(L_1-L_2) \frac{d\hfil}{dL_1}((\gamma+\log L_1)\theta(L_1)),
\eea
where $\gamma$ is the Euler-Mascheroni constant and $\theta(t)$ the step function.

Before moving on to examples for which neither P1 are P2 true it is important to note that the sum in the Seiberg-Shih relation does not respect the symmetry of the Kac table and so for a given amplitude there are two possible Seiberg-Shih relations that might possess P1 or P2. However, there are cases where both possibilities lead to deviations that do not possess either property P1 or P2. For an example of a deviation that does not possess property P1 consider
\bea
\Delta\C{Z}(1,2;\sigma_1|1,1;\sigma_2)= \ln\left[\frac{-1 + 2{x_1}^2 + 2\sqrt{2} x_1 x_2 + 2{x_2}^2}{2(\tilde{x}_1 + \tilde{x}_2)}\right]
\eea
or, since $\C{Z}(2,2;\sigma_1|1,1;\sigma_2)=\C{Z}(1,2;\sigma_1|1,1;\sigma_2)$,
\bea
\Delta\C{Z}(2,2;\sigma_1|1,1;\sigma_2)= \ln \left[ (-1 + 2{x_1}^2 - 2\sqrt{2} x_1 x_2 + 2{x_2}^2)\right],
\eea
where we have suppressed the $\epsilon$ dependent term. It is clear that since the argument of both expressions is not a homogeneous polynomial in $x$ the dependence on $\mu$ cannot be factored out and placed in a separate term. Hence the IR cutoff cannot be chosen to cancel all dependency on the bulk cosmological constant. For an example of a deviation that does not possess property P2 consider,
\bea
\label{P2examplea}
\Delta\C{Z}(1,2;\sigma_1|1,2;\sigma_2)= \ln\left[\frac{\tilde{x}_1 - \tilde{x}_2}{2(x_1 - x_2)}\right]
\eea
or, since $\C{Z}(2,2;\sigma_1|2,2;\sigma_2)=\C{Z}(1,2;\sigma_1|1,2;\sigma_2)$,
\bea
\label{P2exampleb}
\Delta\C{Z}(2,2;\sigma_1|2,2;\sigma_2)= \ln\left[\frac{(x_1 - x_2)(x_1^2+x_2^2-1)}{2(\tilde{x}_1 - \tilde{x}_2)}\right].
\eea
To take the inverse Laplace transform of these deviations it is easiest to use the integral expression for the amplitudes,
\bea
\label{Zintegralc}
\C{Z}(k,l;\sigma_1|r,s;\sigma_2)= \frac{2\pi^2 }{\sqrt{2}pq} \int^{\infty}_0 \frac{dP}{P} \frac{\cos(2\pi\sigma_1 P)\cos(2\pi\sigma_2 P)\sinh(\frac{2\pi P}{\sqrt{pq}})}{\sinh(2\pi b P)\sinh(\frac{2\pi P}{b})} F_{k,l,r,s}(\frac{2\pi i P}{\sqrt{pq}})\nn\\
\eea
where,
\bea 
F_{k,l,r,s}(z)=\sum^{\lambda_p(k,r)}_{\eta=|k-r|+1,2} \sum^{\lambda_q(l,s)}_{\rho=|l-s|+1,2} \sum^{p-1}_{a=1} \sum^{q-1}_{b=-(q-1)} &&\Bigg[\sin(\frac{\pi t}{p})\sin(\frac{\pi t}{q}) \times \\ 
&&U_{\eta-1}(\cos\frac{\pi t}{p})U_{\rho-1}(\cos\frac{\pi t}{q}) \frac{1}{\cos{\frac{\pi t}{pq}}-\cos{z}}\Bigg] \nn
\eea
and $\lambda_p(k,r)=\mathrm{min}(k+r-1,2p-1-k-r)$. This integral is a generalisation of the integral appearing in \cite{Martinec:2003ka} and \cite{Kutasov:2004fg}; its derivation may be found in Appendix \ref{Cylindercalc} \cite{Gesser:2010fi}. Following \cite{Martinec:2003ka} the inverse Laplace transform may be computed by noting that
\bea
\frac{\pi b \cos(2 \pi P \sigma)}{2 P \sinh(\frac{2\pi P}{b})} = \int^\infty_0 \frac{dl}{l} e^{-M l \cosh(\pi b \sigma)} K_{\frac{2iP}{b}}(M l)
\eea
and then substituting this relation into \eqref{Zintegralc}. The result is
\beq
\C{L}^{-1}[\C{Z}](k,l;L_1|r,s;L_2) \propto \int^{\infty}_0 dP \psi_P(L_1)\psi_P(L_2) \frac{\sinh(\frac{2 \pi P}{\sqrt{pq}}) }{\sinh(2\pi P b)} F_{k,l,r,s}(\frac{2\pi i P}{\sqrt{pq}})
\eeq
where $\psi_P(L) = \sqrt{P \sinh(2 \pi P/b)} K_{2iP/b}(L)$. By noting that the integrand is symmetric, the integral may be computed by closing the integral along the entire real axis by a semicircular contour. Since the Seiberg-Shih deviation is merely a linear combination of amplitudes its double inverse Laplace transform with respect to each boundary cosmological constant also exists and is given by
\beq
\C{L}^{-1}[\Delta\C{Z}](k,l;L_1|r,s;L_2) \propto \int^{\infty}_0 dP \psi_P(L_1)\psi_P(L_2) \frac{\sinh(\frac{2 \pi P}{\sqrt{pq}}) }{\sinh(2\pi P b)} \Delta F_{k,l,r,s}(\frac{2\pi i P}{\sqrt{pq}})
\eeq
where
\beq
\Delta F_{k,l,r,s}(z) \equiv F_{k,l,r,s}(z) - U_{k-1}(\cos qz) U_{l-1}(\cos pz) F_{1,1,r,s}(z)
\eeq
and we have assumed that the Seiberg-Shih transformation has been applied to the $(k,l)$ boundary. An important property of $\Delta F_{k,l,r,s}(z)$ is that it is an entire function and therefore the only contributions to the integral comes from the poles at $P = \frac{in}{2b}$ where $n \in \BB{Z}$. It is now easy to show P1 and P2 are not satisfied for \eqref{P2examplea} and \eqref{P2exampleb}; in both cases there exist poles when $P = \frac{in}{2b}$ and therefore when the integral contour is closed around them it will result in a sum of terms containing Bessel functions of the form $K_\frac{4n}{3}(L)$ which have global support in $L$.

We now want to motivate the idea that P3 is a reasonable extension of the criteria for throwing out terms that spoil the Seiberg-Shih relations (and indeed is the most generous we can think of). Although for the moment we use the cylinder amplitudes as a motivation for introducing P3 we will later analyse the case of the disc-with-handle amplitude more carefully.

Consider the integral representation of the amplitude \eqref{Zintegralc}; it may be computed by extending the region of integration to the entire real line and then splitting the integrand up into terms for which the contour may be closed in either the upper or lower half plane. Assuming that $\sigma_1 > \sigma_2$, then upon substituting in \eqref{Fresidues} and \eqref{Gresidues} this results in the expression,
\bea
\label{vExplicitZ}
&&\C{Z}(k,l;\sigma_1|r,s;\sigma_2) \propto F_{k,l,r,s}(0)(\frac{1}{2\epsilon}-\pi \sigma_1)\frac{1}{\sqrt{pq}} + \\
&&\sum_{\substack{n=1\\ n\neq 0modp\\ n\neq 0modq }}\frac{4pq}{n} U_{k-1}(\cos\frac{\pi n}{p})U_{r-1}(\cos\frac{\pi n}{p}) U_{l-1}(\cos\frac{\pi n}{q})U_{s-1}(\cos\frac{\pi n}{q}) e^{\frac{\pi n \sigma_1}{\sqrt{pq}}} \cosh \frac{\pi n \sigma_2}{\sqrt{pq}} + \nn \\
&&\sum^{\infty}_{\substack{n=1\\ n\neq 0\mod q}}  \frac{2}{n} (-1)^n F_{k,l,r,s}(\frac{\pi n}{q})\frac{\sin(\pi n /q)}{\sin(\pi p n /q)} e^{-\frac{\pi p n \sigma_1}{\sqrt{pq}}} \cosh(\pi b n \sigma_2) +\nn\\
&&\sum^{\infty}_{\substack{n=1\\ n\neq 0\mod p}} \frac{2}{n} (-1)^n  F_{k,l,r,s}(\frac{\pi n}{p})\frac{\sin(\pi n /p)}{\sin(\pi q n /p)}e^{-\frac{\pi q n \sigma_1}{\sqrt{pq}}} \cosh (\pi n \sigma_2/b) + \nn \\
&&\sum^{\infty}_{n=1}  \frac{2}{npq} (-1)^{n(p+q+1)} F_{k,l,r,s}(\pi n) e^{-\frac{\pi p q n \sigma_1}{\sqrt{pq}}} \cosh \frac{\pi p q n \sigma_2}{\sqrt{pq}}. \nn
\eea
The above expression may be understood as representing a cylinder amplitude as a sum over disc amplitudes with various local or boundary operators inserted. If we compute the deviation using this expression the first sum cancels while the others remain with the replacement $F_{k,l,r,s} \rightarrow \Delta F_{k,l,r,s}$. The sum in which each term is proportional to $\cosh(\pi b n \sigma_2)$ arises from the poles in \eqref{Zintegralc} due to the factor of $\sinh(\frac{2\pi P}{b})$ in the denominator; it is precisely these terms which vanish \footnote{Actually they have point-like support if we were being more careful.} when we take the double inverse Laplace transform with respect to the boundary cosmological constant. These terms have the interpretation of some form of descendant dual boundary length operator inserted on the boundary of a disc. On the other hand, the sum in which each term is proportional to $\cosh(\pi n \sigma_2/b)$ arises from the poles in \eqref{Zintegralc} due to the factor of $\sinh(2\pi b P)$ in the denominator; it is these terms which, under a double inverse Laplace transform with respect to the boundary cosmological constant, become the terms proportional to Bessel functions of the form $K_\frac{4n}{3}(L_2)$. Similarly, if we were instead to take the double inverse Laplace transform with respect to the dual boundary cosmological constant, the terms proportional to $\cosh(\pi n \sigma_2/b)$ would vanish and the terms proportional to $\cosh(\pi n b \sigma_2)$ would become terms proportional to $K_\frac{3n}{4}(L_2)$. We therefore see that the deviation computed from \eqref{vExplicitZ} can be understood as a sum of terms which vanish under a double inverse Laplace transform with respect to the boundary cosmological constant or its dual.


\section{Testing the Seiberg-Shih relations using a Matrix Model}
\label{MatrixSec}
In the last section we argued that the deviations for all cylinder amplitudes have property P3. Obviously it would be interesting to know whether this is true for deviations of any amplitude. Unfortunately amplitudes more complicated than the disc and cylinder have never been computed using the Liouville approach. An alternative is to use the matrix model formulation of minimal string theory in which the computation of amplitudes with arbitrary numbers of boundaries and handles is straight-forward. 

The partition function of a hermitian $N$-matrix model can be interpreted as generating graphs with vertices carrying $N$ labels with weighting determined by the action of the model. By tuning the parameters that appear in the action we may reach a second order phase transition which will be described by a conformal field theory. By studying the critical exponents, the CFT associated with the phase transition can be identified with Liouville gravity coupled to a minimal model. The disadvantage of using the matrix model is that it gives the amplitude in a fully integrated and summed form and so it is hard to understand the structure of the amplitude in terms of continuum concepts such as states circulating in a loop. Furthermore, generally it is obscure how the graph labels map to the conformal field theory degrees of freedom and even more importantly, the matrix model appears to not contain all the boundary states present in the minimal string. Indeed, this was used as evidence in \cite{Seiberg:2003nm} in support of the Seiberg-Shih relation. However, there are special cases where the relation is manifest; the $(3,4)$ minimal string admits a formulation as a matrix model in which it is easy to construct all boundary states. This is the model we will study in this section. In recent work of \cite{Ishiki:2010wb,Bourgine:2010ja} a number of other boundary states of the $(p,q)$ minimal string have been constructed in the matrix model formulation and it would be interesting to extend our results to those cases.

The $(3,4)$ minimal model coupled to Liouville gravity may be thought of as the CFT that describes the critical point of an Ising model on a random lattice. It is easy to formulate a 2-matrix model in which the weighting of the graphs allows the two labels to be interpreted as the spin degrees of freedom in the Ising model with a coupling dependent on the coupling in the matrix model action. Such a matrix model is
\bea
\label{2MMZ}
\Z=\int dM_1\,dM_2\exp\left(-\frac{N}{g}\,\Tr \left(- cM_1 M_2+\half(M_1^2+M_2^2)+\third(M_1^3+M_2^3)\right)\right)\label{ZXY}
\eea
The Feynman rules for this theory generate graphs of coordination number three with each vertex generated by either $\third M_1^3$ or $\third M_2^3$. Such a graph may be interpreted as a discrete surface composed of triangles with each vertex at the centre of each triangle and with label spin $+$ or $-$ depending on if the vertex is generated by $M_1$ or $M_2$. The coupling between neighbouring spins is controlled by the parameter $c$ whereas $g$ controls the cost associated with adding more vertices to the graph. It is $g$ and $c$ we use to tune the matrix model to its critical point which for our matrix model we do by setting,
\bea
\label{gscale}
g = g_c(1-a^2 \eta \mu), \qquad c = c_c,
\eea
where $\eta$ is some constant which can be determined by comparing with the Liouville theory and the critical point is achieved by letting $a \rightarrow 0$\footnote{The constants $g_c$ and $c_c$ have the well known values $g_c=10c_c^3$ and $c_c = \frac{1}{27}({2\sqrt{7}-1})$.}. We will often refer to this limit at the scaling or continuum limit.

The conformally invariant boundary conditions of minimal CFTs are in one-to-one correspondence with the primary fields of the theory. For a CFT that describes the critical point of some discrete model the boundary conditions can sometimes be understood as universality classes of boundary conditions in the discrete model. As we approach the critical point of a discrete model all boundary conditions in a given class will flow to the same boundary condition in the CFT. In particular for the $(3,4)$ minimal string its three conformal boundary conditions are the continuum limit of the discrete configurations of boundary spins consisting of all spin $+$, all spin $-$, or free spins. It is these boundary conditions we want to implement in the matrix model.

The usual way of inserting a boundary in the matrix model is to compute the resolvent,
\beq W_{M_1}(x)=\avg{\frac{1}{N}\Tr\frac{1}{x-M_1}} \equiv \frac{1}{N}\sum^{\infty}_{n=0} \frac{\avg{\Tr M^n}}{x^{n+1}}.\eeq
The graphs that contribute to each quantity $\avg{\Tr M^n}$ correspond to triangulated surfaces with a boundary of length $n$ composed of vertices generated by $M_1$. The function $W_{M_1}$ is a generating function for such quantities. We will adopt the following notation for more general quantities,
\bea
\label{genresolve}
\lefteqn{W^{Q_1}{}_{F_1}{}^{Q_2}{}_{F_2\ldots;}{}^{Q_3}{}_{G_1}{}^{Q_4}{}_{G_2\ldots;\ldots}(x_1,x_2,\ldots;y_1,y_2,\ldots;\ldots)=}\nn\\
&&\avg{\frac{1}{N}\Tr\Big(\frac{Q_1(M_1,M_2)}{x_1-F_1(M_1,M_2)}\frac{Q_2(M_1,M_2)}{x_2-F_2(M_1,M_2)}\dots\right)\ldots \\ 
&&\frac{1}{N}\Tr\left((\frac{Q_3(M_1,M_2)}{y_1-G_1(M_1,M_2)}\frac{Q_4(M_1,M_2)}{y_2-G_2(M_1,M_2)}\dots\Big)\ldots}\nn
\eea
By tuning the $x_i, y_i, \ldots$ as we take the matrix model to its critical point we can extract continuum quantities corresponding to amplitudes with macroscopic boundaries. By way of example, for $W_{M_1}(x)$ we use \eqref{gscale} and set,
\bea
\label{xscale}
x = x_c(1-a^{d} \kappa \mu_B)
\eea
where $d$ is chosen to produce a non-trivial limit and $\kappa$ is chosen to agree with Liouville theory. As we let $a \rightarrow 0$, $W_{M_1}$ will have an expansion in powers of $a$ \footnote{When taking the scaling limit in the remainder of the paper we will choose $\eta = 1$ and $\kappa = \sqrt{g_c/c_c}$. It should be kept in mind though that by changing $\eta$ and $\kappa$ we can renormalise $\mu_B$ and $\mu$.},
\bea
W_{M_1}(x_c(1-a^{d} \kappa \mu_B)) = \sum_i a^{d_i} W_i(\mu_B,\mu) + a^{d_W} \tW_{M_1}(\mu_B,\mu) + h.o.t \quad.
\eea
In this expression $W_i(\mu_B,\mu)$ is analytic in both arguments whereas $\tW_{M_1}$ is defined as being the first term non-analytic in $\mu_B$ and $\mu$. It is $\tW_{M_1}$ that corresponds to the continuum quantity and so in this case is the partition function of the continuum theory defined on a surface with a finite sized boundary with the boundary condition determined by which universality class the discrete quantities $\avg{\Tr M^n}$ belong to. This is slightly complicated by the fact that the resolvent actually corresponds to a boundary with a marked point and therefore must be integrated with respect to the boundary cosmological constant before comparing to Liouville theory. This motivates introducing the integrated quantity,
\bea
\omega({\mu_B}_1,...,{\mu_B}_n,\mu) \equiv \int \prod_i d{\mu_B}_i \tW({\mu_B}_1,...,{\mu_B}_n,\mu)
\eea
where we have allowed for more general amplitudes which have more than one boundary. The generalisation to amplitudes of the form \eqref{genresolve} is obvious.

For the matrix model \eqref{ZXY} it is easy to construct resolvents which generate boundary conditions that flow to the three different boundary conditions in the CFT:
\bea
\label{boundaryResolve}
W_{M_1}(x)&=&\avg{\frac{1}{N}\Tr\frac{1}{x-M_1}}, \nn \\
W_{M_2}(x)&=&\avg{\frac{1}{N}\Tr\frac{1}{x-M_2}},  \\
W_{M_1+M_2}(x)&=&\avg{\frac{1}{N}\Tr\frac{1}{x-(M_1+M_2)}}. \nn
\eea
The first two resolvents $W_{M_i}$ generate surfaces on which only $M_i$ vertices appear on the boundary. Since the $M_i$ vertices map directly to the spin degrees of freedom these flow to the fixed spin boundary conditions in the continuum limit. The final resolvent generates boundaries in which both types of vertices appear with equal weighting. This will therefore flow to the free spin boundary condition. We can parameterise these resolvents in the following way,
\bea
\label{genXresolvent}
W_X(x) = \avg{\frac{1}{N}\Tr\frac{1}{x-X}} \qquad X = \frac{(1 - \alpha)}{2}M_1-\frac{(1 + \alpha)}{2}M_2-(1 + c) \frac{\alpha}{2}.
\eea
This generates graphs with weighting corresponding to the Ising model on a random lattice but with a boundary magnetic field, to which the parameter $\alpha$ is related. Clearly, we can choose $\alpha=\pm 1$ to reproduce the resolvents $W_{M_i}$. By taking $\alpha$ to infinity we may also obtain the resolvent $W_{M_1+M_2}$. One method to compute the resolvent $W_X$ is to make the following change of variable in \eqref{ZXY},
\bea
\label{SXdef}
S = M_1+M_2+1+c, \qquad
X = \frac{(1 - \alpha)}{2}M_1-\frac{(1 + \alpha)}{2}M_2-(1 + c) \frac{\alpha}{2}.
\eea
The partition function in these new variables takes the form,
\bea
\label{ZSX}
\mathcal{Z}=\int [dS dX] \exp{\left[-\frac{N}{g}\Tr\left(X^2 S+\alpha XS^2+V(S)\right)\right]}
\eea
where,
\beq 
\label{VSX}
V(S)=\frac{1}{12}(1+3\alpha^2)S^3-\frac{c}{2}S^2+\quarter(3c-1)(1+c)S.
\eeq
Generic values of $\alpha$ (i.e. the Ising model on a random lattice in the presence of a boundary magnetic field) were investigated using matrix model techniques in \cite{Carroll:1995qd}, in which the disc amplitudes were calculated. The method employed in \cite{Carroll:1995qd} to solve the matrix model is combinatorial and so is not easily generalised to compute more complicated amplitudes and so we will not employ it here. However, it is clear that if we are only interested in amplitudes in which all the boundary conditions are of the free spin form we may set $\alpha = 0$ in \eqref{ZSX} to obtain the $O(1)$ model whose solution is well known. Similarly if we are only interested in amplitudes in which all the boundary conditions are of the fixed spin form we may simply use \eqref{2MMZ}. This is in fact the case for most of the results we require for the remainder of this section and so they can be obtained form the literature  \cite{Eynard:2002kg}\cite{Eynard:1995nv}. For more complicated amplitudes involving both fixed and free spin boundary conditions it is necessary to solve the matrix model \eqref{ZSX}; this may be done using loop equations and we give details of this in Appendix \ref{MMloops}\footnote{We have, as a cross-check, computed all the amplitudes in this paper using the loop equation method detailed in the appendix.}. Although we may look up many of the results in the following section in the literature, we will, in order to discuss all the matrix model results in a common language, phrase the discussion as if we had computed the various resolvents from \eqref{ZSX} using the loop equations.

To take the scaling limit of $W_X$ at $\alpha = -1$ we may use the results of \cite{Eynard:2002kg} or the results in Appendix \ref{MMloops}. The critical value of $x$ is $x=-c$ and the solution to the loop equation gives, in the scaling limit,
\beq \tW_{M_1}(\mu_B,\mu)=-\frac{1}{2^{{\frac{5}{3}}}5^\frac{1}{3}c}\left(\left(\mu_B+\sqrt{\mu_B^2-\mu}\right)^\frac{4}{3}+ \left(\mu_B-\sqrt{\mu_B^2-\mu}\right)^\frac{4}{3}\right)\label{WX}.\eeq
For ${W_S}^{(0)}$ we may use \cite{Eynard:1995nv} or solve \eqref{WSeqn}. The critical value of $x$ is at $x=0$ and computing the scaling limit of ${W_S}^{(0)}$ we get,
\beq
\tW_{M_1+M_2}(\mu_B,\mu)=-\frac{1}{2^{{\frac{5}{3}}}5^\frac{1}{3}c}\left(\left(\frac{\mu_B}{\sqrt{2}}+\sqrt{\frac{\mu_B^2}{2}-\mu}\right)^\frac{4}{3}+ \left(\frac{\mu_B}{\sqrt{2}}-\sqrt{\frac{\mu_B^2}{2}-\mu}\right)^\frac{4}{3}\right).\label{WS}
\eeq
Here we run into one final technicality. For a resolvent that produces graphs with some configuration of spins on the boundary $W_{F(M_1,M_2)}$, there exist an entire family of resolvents which produce the same configuration of boundary spins but with different overall weighting
\beq
W_{\frac{1}{\lambda} F(M_1,M_2)}(x) = \lambda W_{F(M_1,M_2)}(\lambda x),
\eeq
where $\lambda$ is a constant. Changing $\lambda$ has the affect of renormalising $\mu_B$, however we want the same normalisation for all continuum quantities and so this fixes $\lambda$ to be $\sqrt{2}$ for $W_{M_1+M_2}$. Therefore using \eqref{WS} and \eqref{WX}, we have that 
\beq
\omega_{(M_1+M_2)/\sqrt{2}}\,(\mu_B,\mu)=\sqrt{2}\omega_{M_1}(\mu_B,\mu).
\eeq
This is precisely the relation between these two disc amplitudes found in the continuum calculation.

One may wonder what the scaling limit of ${W_{X}}^{(0)}$ is for generic values of $\alpha$; given that there are only a finite number of Cardy states the expectation is that there should be no new non-trivial scaling limits. For generic $\alpha$ the critical values of $x$ occur at $\frac{2c}{ \alpha-1}$ and $\frac{2c}{ \alpha+1}$. Taking the scaling limit, for both critical values of $x$, gives
\beq
\omega_{-X/\alpha}(\mu_B,\mu)=\omega_{M_1}(\mu_B,\mu)\label{WXa}
\eeq
which confirms our expectation. We have therefore demonstrated that we can build all the boundary states of the $(3,4)$ minimal string. This will be confirmed further by reproducing the cylinder amplitudes shortly. The nature of the scaling limit when $\alpha = 0$ will be addressed in the Conclusion \eqref{conclusion}.

The one case in which we are interested which cannot be found in the literature is $W_{S;X}$. This may be obtained by the method of loop equations and we give the details in Appendix \ref{MMloops}. The computation results in
\bea
\label{WSXeqn}
\omega_{S;X}({\mu_B}_1,{\mu_B}_2,\mu)=-\frac{1}{10 c_c^2} \log(-1 + 2 z_1^2 + 2 \sqrt{2} z_1 z_2 + 2 z_2^2)
\eea
for the scaling form of $W_{S;X}$, which is in exact agreement with the Liouville result (up to renormalisations of $\mu_B$ and $\mu$). Of course the other cylinder amplitudes may computed in this matrix model or using \cite{Eynard:2002kg}\cite{Eynard:1995nv} and we find the scaling forms to be,
\bea
&&\omega_{X;X}({\mu_B}_1,{\mu_B}_2,\mu)=\frac{1}{10 c_c^2} \ln \frac{z_1-z_2}{x_1-x_2},\\
&&\omega_{X;Y}({\mu_B}_1,{\mu_B}_2,\mu)= -\frac{1}{10 c_c^2} \ln z_1+z_2,\\
&&\omega_{S;S}({\mu_B}_1,{\mu_B}_2,\mu)=\frac{1}{10 c_c^2} \ln \frac{z_1-z_2}{(z_1+z_2)(x_1-x_2)}, 
\eea
again in agreement with the Liouville calculations. The fact that these amplitudes agree with the Liouville calculations \emph{including} terms that were classified in \cite{Kutasov:2004fg} as non-universal suggests they really should have a physical meaning (and shouldn't be thrown away). This issue was raised in \cite{Belavin:2010pj}.

In order to test if P3 is a property of all deviations we need to examine more complicated amplitudes. The $1/N$ correction to the disc amplitude is the amplitude for a disc-with-handle. These computations have already been done in \cite{Eynard:1995nv} and \cite{Eynard:2002kg}, the result being,
\bea
\label{DWHresolv}
{\omega_{M_1}^{(1)}(\mu_B,\mu)}={\omega_{M_2}^{(1)}(\mu_B,\mu)}&=&-\frac{1}{2^\frac{1}{3} 5^\frac{2}{3} 36c_c }\frac{z (7 - 24 z^2 + 48 z^4)}{(-1 + 4 z^2)^3}\mu^{-\frac{13}{6}},\\
{\omega_{(M_1+M_2)/\sqrt{2}}^{(1)}(\mu_B,\mu)}&=&-\frac{\sqrt{2}}{2^\frac{1}{3} 5^\frac{2}{3} 36c_c }\frac{z (5 - 24 z^2 + 48 z^4)}{(-1 + 4 z^2)^3} \mu^{-\frac{13}{6}}.
\eea
If we now compute the deviations we find,
\bea
\label{DeltaHandle}
\Delta{\omega_{M_2}^{(1)}(\mu_B,\mu)} &=& 0,\\
\Delta{\omega_{(M_1+M_2)/\sqrt{2}}^{(1)}(\mu_B,\mu)} &=& \frac{2^\frac{1}{6} 4}{5^\frac{2}{3} 9 c_c} \frac{z (-3 + 41 z^2 - 208 z^4 + 416 z^6 - 512 z^8 + 256 z^{10})}{ (-1 + 2 z)^3 (1 + 2 z)^3 (1 - 16 z^2 + 16 z^4)^3} \mu^{-\frac{7}{6}}. \nn
\eea

We now want to consider if $\Delta{\omega_{(M_1+M_2)/\sqrt{2}}^{(1)}(\mu_B,\mu)}$ possesses P3. First note that if we write
\beq \Delta{\omega_{(M_1+M_2)/\sqrt{2}}^{(1)}(\mu_B,\mu)}=A(x)+B(\tilde {x})\eeq
then in order for $A$ and $B$ to have an interpretation as the contribution of certain geometries to the amplitude then both $A$ and $B$ must be positive for physical values of their arguments, in particular as $x,\tilde{x} \rightarrow \infty$ \footnote{This may seem to contradict the expressions given earlier for cylinder amplitudes e.g \eqref{P2examplea} however in this case some of the contribution from the singular geometries has cancelled between $A$ and $B$; if they are reintroduced then the resulting $A'$ and $B'$ will be positive.}. Furthermore, since
\bea
\Delta{\omega_{(M_1+M_2)/\sqrt{2}}^{(1)}(\mu_B,\mu)} = \frac{2^\frac{-11}{6} }{5^\frac{2}{3} 9 c_c} \frac{1}{z^7}
\sim\frac{1}{x^{7/3}}\sim\frac{1}{\tilde{ x}^{7/4} }
\eea
this means that $A(x)= a x^{-7/3} +o(x^{-7/3})$ or $B(x)=b{x^{-7/4}} +o(x^{-7/4})$ otherwise for sufficiently large $x$ either $A$ or $B$ would be negative. We therefore conclude both the inverse Laplace transforms of $A$ and $B$ with respect to either $x$ or $\tilde{x}$ exist. For $\Delta{\omega_{\frac{1}{\sqrt{2}}(M_1+M_2)}^{(1)}}$ to possess P3 we require that $\C{L}^{-1}_x[A]$ and $\C{L}^{-1}_{\tilde{x}}[B]$ \footnote{We denote in the subscript the variable the inverse Laplace transform is taken with respect to.} have point-like support. However it is a well known theorem that a function with only point-like support may be expressed as a sum of derivatives of $\delta$-functions,
\beq
f(t) = \sum^\infty_{m=1}\sum^\infty_{n=1} a_{n,m} \delta^{(n)}(t - t_{n,m})
\eeq
and such a function will have a Laplace transform of
\beq
\label{inverseLaplace}
\C{L}[f](s) = \sum^\infty_{m=1}\sum^\infty_{n=1} a_{n,m} s^n e^{-s t_{n,m}}.
\eeq
Given that we require that $\C{L}^{-1}_x[A]$ and $\C{L}^{-1}_{\tilde{x}}[B]$ are point-like supported this implies $A(x)$ and $B(x)$ have an asymptotic behaviour as $x \rightarrow \infty$ inconsistent with \eqref{DeltaHandle} and hence the deviation can not possess P3.

\section{The Seiberg-Shih Relation and Local States}
\label{Localstates}
Suppose that the boundaries considered in the previous sections could be expressed as a sum over local operators in the theory, such as,
\beq
\label{localopexp}
\bra{k,l;\sigma_1} \sim \sum^\infty_{n=1} f_n(k,l,\sigma_1) \bra{O_n},
\eeq
where $O_n$ represents some generic local operator and $\sim$ means equivalence up to terms that diverge as $\sigma_1 \rightarrow \infty$. If this were true then the effects of the Seiberg-Shih transformations on a boundary would reduce to studying their effect on the above series. The expansion of a boundary in terms of local operators was investigated in \cite{Moore:1991ir} in which it was used to compute the one and two point correlation numbers of tachyon operators on the sphere for the $(2,2p+1)$ minimal strings. Later it was shown \cite{Anazawa:1997zs} that such a technique could also compute the three point function correctly for the $(p,p+1)$ models.

One technical point raised in \cite{Moore:1991ir} and \cite{Anazawa:1997zs} was that the boundary state could only be expressed as a sum over local operators if we included operators not in the BRST cohomology of \cite{Lian:1992aj}. That the operators in the cohomology of \cite{Lian:1992aj}, which we refer to as LZ-operators, are not all the physical operators of the minimal string is easily verified by noting that the cohomology does not contain the boundary length operator $\oint_{\partial M} dx e^{b\phi}$. Such an operator can be inserted on the disc by differentiating with respect to the boundary cosmological constant, giving the result
\beq
\avg{\oint_{\partial M} dx e^{b\phi}} =\C{A}(p,q) U_{k-1}\left(\cos \frac{\pi q}{p} \right) U_{l-1}\left( \cos \frac{\pi p}{q} \right)\cosh\left(\frac{\pi q \sigma}{\sqrt{pq}}\right),
\eeq
where $\C{A}(p,q)$ is a constant of proportionality that only depends on $p$ and $q$. Indeed there exists a whole family of such operators with Liouville charge $b\frac{p+q-n q}{2p}$ where $n \in \BB{Z}^+$ which we will denote by $\C{U}_{nq}$. 

In order to compute the coefficients $f_n(k,l,\sigma_1)$ appearing in \eqref{localopexp} we need to know the amplitude for any local operator inserted in a disc with a $(k,l)$ FZZT boundary condition. For the tachyon operators the amplitude is \eqref{tachyonDisk}. For the non-tachyon operators we have not computed the exact amplitude using Liouville theory, however it is easy to find the contribution to the amplitude from the Liouville sector in the semi-classical approximation by solving the Wheeler-DeWitt equation, combined with the matter sector we get \footnote{In fact the semi-classical approximation has been shown to give the exact answer in the case of LZ-operators inserted in the disc; we are going to assume this is also the case for non-LZ operators.}
\beq
\braket{\C{U}_{nq}}{k,l;\sigma} \propto \cosh \left(\frac{\pi n q \sigma}{\sqrt{pq}} \right).
\eeq
Finally, if we apply the Liouville duality transformation to the above expression we get
\beq
\braket{\C{U}_{np}}{k,l;\sigma}_{\mathrm{dual}} \propto \cosh \left(\frac{\pi n p \sigma}{\sqrt{pq}} \right),
\eeq
where we have introduced the operators $\C{U}_{np}$ which have Liouville charge $b\frac{p+q-n p}{2p}$. The expansion \eqref{localopexp} can now be given a more concrete form,
\beq
\label{localopexp2}
\bra{k,l;\sigma_1} \sim \sum^\infty_{n=1} f_n(k,l,\sigma_1) \bra{\C{U}_n}.
\eeq

Having introduced the necessary technical results we now want to prove the following claim,
\newline
\newline
{\emph{
If all the FZZT branes in the $(p,q)$ minimal string can be replaced by local operators then the deviation from the Seiberg-Shih relations is caused only by the non-LZ operators.
}}
\newline
\newline
Consider the cylinder amplitude $\C{Z}(k,l;\sigma_1|r,s;\sigma_2)$. If we replace one of the boundaries by a sum of local operators using \eqref{localopexp2} then we get,
\beq
\label{opexpinCyl}
\C{Z}(k,l;\sigma_1|r,s;\sigma_2) = \sum_{n=1}   f_n(k,l,\sigma_1) \braket{\C{U}_n}{r,s;\sigma_2}.
\eeq
By comparison of \eqref{vExplicitZ} with \eqref{opexpinCyl}, we find for the coefficient of the LZ operators,
\bea
\label{fkleqn}
f_n(k,l;\sigma_1) = -\frac{4pq A_C}{A_D(n)}\frac{1}{n} U_{k-1}(\cos\frac{\pi n}{p}) U_{l-1}(\cos\frac{\pi n}{q})\mu^{-\frac{n}{2p}}e^{-\frac{\pi n \sigma_1}{\sqrt{pq}}} \equiv \tilde{f}_n(k,l)e^{-\frac{\pi n \sigma_1}{\sqrt{pq}}},
\eea
where $A_C$ is just a numerical constant and $n$ is not a multiple of $p$ or $q$. Furthermore, note that the LZ operator coefficients satisfy
\beq
f_n(k,l;\sigma_1) = \sum^{k-1}_{a=-(k-1),2} \sum^{l-1}_{b=-(l-1),2} \tilde{f}_n(1,1)e^{-\frac{\pi n}{\sqrt{pq}}\left( \sigma_1 + i\frac{qb+pa}{\sqrt{pq}}
\right).} 
\eeq
This shows that if all boundary states admit an expansion in terms of local operators then the Seiberg-Shih relations transform the coefficients of the LZ-operators correctly. Hence any deviation from the Seiberg-Shih relations must come from the non-LZ operators.
However, it is the non-LZ operators one would expect to give deviations compatible with P3 as they correspond to boundary operators and their duals. One should then be suspicious that in fact the boundary states may not be expressed as a sum over local operators. We will now see this suspicion is borne out by examination of the cylinder amplitudes. 

If we note that for cylinder amplitudes, $\C{Z}(k,l;\sigma_1|r,s;\sigma_2) = \C{Z}(k,l;\sigma_2|r,s;\sigma_1)$ then together with \eqref{localopexp2}, this implies,
\beq
\sum^\infty_{n=1} f_n(k,l,\sigma_1) \braket{\C{U}_n}{r,s;\sigma_2} = \sum^\infty_{n=1} f_n(r,s,\sigma_1) \braket{\C{U}_n}{k,l;\sigma_2}.
\eeq
Since $\braket{\C{U}_n}{r,s;\sigma} = A^{(r,s)}_n g_n(\sigma)$ where $g_n$ form a set of linearly independent functions then this implies
\beq
f_n(k,l,\sigma) = A^{(k,l)}_n h_n(\sigma),
\eeq
where $h_n$ is some function. We conclude that if all states admit an expansion of the form \eqref{localopexp2} then cylinder amplitudes can be expressed as,
\beq
\label{factorform}
\C{Z}(k,l;\sigma_1|r,s;\sigma_2) = \sum^\infty_{n=1} A^{(k,l)}_n A^{(r,s)}_n h_n(\sigma_1) g_n(\sigma_2).
\eeq
We now want to see if the $\sigma$-independent coefficients of each term in \eqref{vExplicitZ} has the above property. Consider the coefficients appearing in the second sum of \eqref{vExplicitZ},
\bea
F_{k,l,r,s}(\frac{\pi n}{q}) &&\frac{\sin(\pi n /q)}{\sin(\pi p n/q)} = \\
&&\sum^{\lambda_p(k,r)}_{\eta=|k-r|+1,2} \sum^{\lambda_q(l,s)}_{\phi=|l-s|+1,2}
 (-1)^{\eta n} U_{\phi-1}\left( \cos \frac{\pi p n \phi}{q} \right) \left(\Theta(q\eta-p \phi)-\frac{\eta}{p}\right), \nn
\eea
where $\lambda_p(k,r)=\mathrm{min}(k+r-1,2p-1-k-r)$ and $n\neq 0 \mod q$. We want to know whether this can be written in the form $A^{(k,l)}_n A^{(r,s)}_n$. That this is not true in general is shown by considering the case $(p,q)=(4,5)$, for which we have,
\bea
&&F_{2,2,2,1}(\frac{\pi n}{q}) \frac{\sin(\pi n /q)}{\sin(\pi p n/q)}=0,  \\
&&F_{1,1,2,1}(\frac{\pi n}{q}) \frac{\sin(\pi n /q)}{\sin(\pi p n/q)}=\half, 
\eea
which is clearly incompatible with the factorisation required \eqref{factorform}. This lack of factorisation appears to be generic; for instance if we consider the $(3,4)$ model we find that there is not even a subspace of states which admit an expansion in terms of local operators. This can be seen by considering the matrix,
\bea\label{nonnulldet}
&&F_{k,l,r,s}(\frac{\pi n}{q}) \frac{\sin(\pi n /q)}{\sin(\pi p n/q)}=
\left( \begin{array}{ccc}
\frac{2(-1)^n}{3} & \third & \frac{2}{3}\cos \frac{3 \pi n}{4} \\
\third & \frac{2(-1)^n}{3} & \frac{2}{3}(-1)^{n+1}\cos \frac{3 \pi n}{4} \\
\frac{2}{3}\cos \frac{3 \pi n}{4} & \frac{2}{3}(-1)^{n+1}\cos \frac{3 \pi n}{4} & \frac{1}{3}(-1)^n(3 - 4 \cos^2\frac{3 n \pi}{4}) \end{array} \right)\nn \\
\eea
where the columns and rows correspond to $(k,l) = (1,1), (2,1), (1,2)$.  If there exist $A^{(k,l)}_n$ and $A^{(r,s)}_n$ such that $F_{k,l,r,s}(\frac{\pi n}{q}) =A^{(k,l)}_n A^{(r,s)}_n$, then the determinant of the matrix in \eqref{nonnulldet} must be zero which is not the case.

The same conclusions can be reached by studying the other coefficients associated to the non-LZ operators in \eqref{vExplicitZ}. This leaves us in an awkward position; there is no subspace of states that can be expanded in terms of local operators unless we allow the subspace to be one-dimensional (in which case why choose a particular boundary as being the one that can be expanded in terms of local operators), yet from the results of \cite{Moore:1991ir} and \cite{Anazawa:1997zs} it is clear information about insertions of operators can be extracted from loops. It is worth noting that from the examination of cylinder amplitudes in which one of the boundaries is a ZZ brane it appears that ZZ branes do not couple to LZ states \cite{Gesser:2010fi,Ambjorn:2007ge,Ambjorn:2007xe}.

An interesting observation is that all correlation numbers were extracted from loops with fixed spin boundary conditions. We have found further evidence that correlation numbers can be extracted from the fixed spin boundary by computing the 1-point correlation numbers on the torus starting form the disc-with-handle amplitude in the case of the $(2,5)$ model. We have found exact agreement with the computation done in \cite{Belavin:2010pj}. A trick that makes this computation easy is that all amplitudes for the minimal string are easily expressible in terms of $z$, which is the uniformising parameter of the auxiliary Reimann surface \cite{Seiberg:2003nm} or equivalently spectral curve. If we write these amplitudes in terms of a new variable, $w$, defined by $z = \half(w+w^{-1})$ we may easily compute the large $w$ expansion of the amplitudes. Since $z = \cosh \frac{\pi \sigma}{\sqrt{pq}}$, the large $w$ expansion will be an expansion in terms of the functions $e^{\frac{\pi  \sigma}{\sqrt{pq}}}$ which is exactly what is required to compute the correlation numbers. Explicitly, if we have an amplitude $\braket{1,1;\sigma_1}{X}$ then
\bea
\braket{1,1;\sigma_1}{X} = \sum^{\infty}_{n = 1} A_n w^{-n} = \sum^{\infty}_{n = 1} A_n e^{-\frac{\pi n\sigma_1}{\sqrt{pq}}},
\eea
but we also have
\bea
\braket{1,1;\sigma_1}{X} = \sum^\infty_{n=1}\tilde{f}_n(1,1)  e^{-\frac{\pi n\sigma_1}{\sqrt{pq}}}\braket{\C{U}_n}{X}
\eea
giving,
\bea
\braket{\C{U}_n}{X} = \frac{A_n}{\tilde{f}_n(1,1) }.
\eea
The reproduction of the torus correlation numbers in Appendix \ref{torusCorrelationNo} is evidence that this procedure is correct and that the fixed spin amplitude can be used to compute correlation numbers of local operators. Furthermore, we can obtain evidence that the non-fixed spin boundary states do not admit an expansion in terms of local operators by applying the above procedure to the disc-with-handle amplitudes we computed for the $(3,4)$ model. We have the expansions,
\bea
{\omega_{M_1}^{(1)}(\mu_B,\mu)}&\sim&\frac{1}{2^\frac{1}{3} 5^\frac{2}{3} c \Lambda^\frac{7}{6}}\left[ \frac{1}{24 w} + \frac{1}{72 w^5} - \frac{5}{72 w^7} + \frac{17}{72 w^{11}} + \C{O}(w^{-13})\right] \\
{\omega_{(M_1+M_2)/\sqrt{2}}^{(1)}(\mu_B,\mu)}&\sim&\frac{1}{2^\frac{1}{3} 5^\frac{2}{3} c \Lambda^\frac{7}{6}}\left[ \frac{1}{24 w} - \frac{1}{72 w^5} - \frac{1}{72 w^7} + \frac{7}{72 w^{11}}  +\C{O}(w^{-13}) \right]
\eea
If the expansion in terms of local operators is valid for both these boundary states then we would expect the coefficients of each term in the large $w$ expansion to be related by a factor of $\sqrt{2}$ as can been from \eqref{fkleqn}; this is clearly not the case. The implication of these results is that the fixed spin boundary condition is special and is the only one that admits an expansion in terms of local operators.

\section{Conclusions}
\label{conclusion}
In this paper we gave examined the Seiberg-Shih relations \eqref{SS1} for cylinders and discs-with-handles. For cylinders we reproduced the result of \cite{Kutasov:2004fg}\cite{Gesser:2010fi} that the identification of FZZT branes is naively spoiled by terms arising from the gravitational sector of the theory and investigated the claim in \cite{Kutasov:2004fg} that all such terms are non-universal. We found that the terms spoiling the identification cannot always be associated with degenerate geometries of the worldsheet as other geometries which are degenerate in a dual sense are also present. This lead us to conjecture that the terms spoiling the FZZT brane identification could always be interpreted as arising from degenerate and dual degenerate geometries. We checked this conjecture by computing the disc-with-handle amplitudes for free and fixed spin boundary states in the $(3,4)$ minimal string using matrix model techniques. We found that the deviation from the Seiberg-Shih relations in this case could not  be written in the way we conjectured. Given that our conjecture was a very generous interpretation of what terms might be unphysical we conclude this is strong evidence that the FZZT brane identification conjectured in \cite{Seiberg:2003nm} does not hold at all levels of perturbation theory.

We also considered an alternative approach to testing the Seiberg-Shih relations by expanding boundaries in terms of local operators. If such an expansion were possible we showed that it would lead to deviations from the Seiberg Shih relations consistent with our conjecture. We then gave explicit examples for cylinder amplitudes where a local operator expansion of boundary states fails. This lead us to a paradox; how could correlation numbers of local operators be extracted from boundary states in \cite{Moore:1991ir} \cite{Anazawa:1997zs} if such boundary states cannot be expanded as local operators? To investigate this we computed the disc-with-handle amplitude in the $(2,5)$ model using matrix model methods and then showed that the fixed spin boundary condition yielded an expansion from which correlation numbers in agreement with the results of \cite{Belavin:2010pj} could be extracted. This lead us to conjecture that the fixed spin boundaries are special in that they allow an expansion in terms of local operators.

There are many obvious generalisations of the work done here, the most obvious being that it would be interesting to extend the above results to the $(p,q)$ case. Since the Liouville techniques are not suited to higher-genus computations such a project would have to be tackled using matrix model methods. The main problem with this approach is that it is difficult to represent all the states for a given minimal string in a matrix model; in the above we chose the Ising model precisely because there was an obvious mapping between the spin degrees of freedom and the matrix fields. However this problem may now be less serious given recent work \cite{Ishiki:2010wb,Bourgine:2010ja} in which many boundary states were constructed in the matrix model formulation. Another strategy for attacking the same problem may be the geometric recursion techniques developed by Eynard et al. in \cite{2008arXiv0811.3531E} as this provides an efficient way of computing many amplitudes for a given matrix model. Such methods
would also allow a relatively easy computation of the cylinder-with-handle amplitudes;
these would be interesting to study as one could then check if quantum corrections destroy
the fact, noted in \cite{Gesser:2010fi,Ambjorn:2007ge,Ambjorn:2007xe}, that FZZT-ZZ cylinder amplitudes are consistent with the Seiberg-
Shih relations.

Another generalisation would be to perform a similar investigation in other non-critical string models. An obvious choice would be non-critical superstrings as they perhaps have more physical relevance in addition to being much better behaved. A second and perhaps less obvious choice of string model would be the causal string theories developed in \cite{Ambjorn:2008gk,Ambjorn:2008hv,Ambjorn:2009fm}. Currently these models have only been solved for a target space of zero dimensions and so such a project would require a generalisation of the models to include worldsheet matter. However, even in the zero dimensional case such causal string models are better behaved and many of the odd features of the usual non-critical string are absent due to the lack of baby-universe production; in particular the gravity sector in these theories seems to be much weaker. Since the terms spoiling the FZZT brane identification are due to the gravity sector (at least for the cylinder amplitudes) might they have an interpretation in terms of baby-universe over production and if so might they be absent in these causal string models?

Finally, perhaps the most obvious omission in this paper is a discussion of the nature of the dual branes. In \cite{Seiberg:2003nm} the two matrix model description of the $(p,q)$ minimal string was considered and it was shown that the continuum limit of the resolvent for one of the matrices was the disc amplitude with a fixed spin boundary. The continuum limit of the resolvent for the other matrix gave the dual brane amplitude. In our matrix model \eqref{ZSX} it is not obvious if we can construct a correlation function which gives the disc with dual brane boundary conditions. In fact it is present in the form,
\beq
W_A(x) = \avg{\frac{1}{N}\Tr\frac{1}{x-(M_1-M_2)}}
\eeq
which may be obtained from the general $X$ resolvent \eqref{genXresolvent} by setting $\alpha = 0$. The resulting loop equation is easily solved as it is quadratic in $x^2$, however the scaling limit is obtained using a different scaling of $x$,
\beq x={x}_c+a^\frac{4}{3}\tilde{\mu}_{B}\sqrt{g_c/c},\label{dualxscale}\eeq
which results in
\beq \tW_{k(M_1-M2)}(\tilde{\mu}_B,\mu)=-\frac{1}{2^{{\frac{5}{3}}}5^\frac{1}{3}c}\left(\left(\tilde{\mu}_B+\sqrt{\tilde{\mu}_B^2-\tilde{\mu}}\right)^\frac{3}{4}+ \left(\tilde{\mu}_B-\sqrt{\tilde{\mu}_B^2-\tilde{\mu}}\right)^\frac{3}{4}\right),\eeq
where $k$ is a numerical constant. If we identify $\tilde{\mu}_B$ in this expression with the dual boundary cosmological constant then this is precisely the amplitude expected for the dual brane obtained from the spin $+$ or spin $-$ boundary state. This explains the $\alpha=0$ possibility we left unexplained at the end of Section \ref{MatrixSec}. We therefore see that the matrix model is able to reproduce all the boundary states found in the $(3,4)$ minimal string apart from the dual of the free spin boundary condition. In \cite{Seiberg:2003nm} this was enough as they claimed that there is only one brane and hence only one dual brane in the theory. However, in light of our results there should exist other dual branes in the theory. Are such states present in the matrix model and if so can they be represented as some form of resolvent? This question and the others outlined we leave to future work.

\acknowledgments
This work is supported by STFC grant ST/G000492/1. MA would like to acknowledge the support of STFC studentship PfPA/S/S/2006/04507. We would also like to thank Shoichi Kawamoto and Simon Wilshin for valuable discussions.
\bibliographystyle{JHEP}


\appendix

\section{Computation of torus 1-point correlation numbers}
\label{torusCorrelationNo}
The disc-with-handle amplitude for the $(2,2p+1)$ minimal string has been computed numerous times using matrix model techniques. For the case of the $(2,5)$ model it takes the form,
\beq
\label{dischandle25}
{\omega_{M}(\mu_B,\mu)}^{(1)} = -\frac{1}{2^\half 3^\frac{3}{4}576 }\frac{1 + 12 z^2}{z^3}\mu^{-\frac{7}{4}}
\eeq
The boundary condition on the disc is the equivalent of our spin up boundary as it is computed using the resolvent of $M$ where $M$ is the only matrix appearing in the matrix model. We will now use this to compute the torus one point correlation numbers, for tachyon operators, by replacing the boundary with a sum of local operators as in \eqref{localopexp2}. That this computation produces results that match exactly the results of \cite{Belavin:2010pj} is evidence that we may replace this boundary by a sum of local operators. Obviously, in order to compare results we have to renormalise the bulk cosmological constant, boundary cosmological constant and $M$ to be consistent with \cite{Belavin:2010pj}. In order to avoid doing this we note that the effect of such a renormalisation will to be to change the amplitude by a factor dependent only on $p$ and $q$ and so will not affect the ratio of correlation numbers. We therefore shall compare the ratio of correlation numbers. Computing the large $w$ expansion of \eqref{dischandle25} we get,
\beq
{\omega_{M}(\mu_B,\mu)}^{(1)}  = 4 \sum^{\infty}_{n=0}  (-1)^{n+1} (n-2)(n+3) w^{-2n-1}
\eeq
where the series converges if $w>1$. We therefore find the torus $1$-point correlation number for tachyon operators in the $(2,5)$ minimal string to be,
\bea
\label{1pointfromloop}
\avg{\C{U}_n}^{(1)} = \frac{A_D}{40 A_C} n \left( \sin\frac{\pi n}{2} \sin\frac{\pi n}{5} \right)^{\half} \Gamma\left(1-\frac{n}{5}\right)\Gamma\left( -\frac{n}{2}\right)(-1)^{\frac{n+1}{2}}(n-5)(n+5) \mu^\frac{n}{4} 
\eea
when $n$ is odd and zero if $n$ is even. The 1-point correlation numbers on the torus for the $(2,2p+1)$ models were computed in \cite{Belavin:2010pj}, giving,
\bea
\avg{\C{U}_n}^{(1)} = \avg{\C{U}_1}^{(1)} (-1)^\frac{n-1}{4}\left(\frac{\sin\frac{\pi n}{2p+1}}{\sin\frac{\pi }{2p+1}} \right)^\half \frac{\Gamma(1-\frac{n}{2p+1})\Gamma(\half)}{\Gamma(1-\frac{1}{2p+1})\Gamma(\frac{1}{n})} \frac{(2p+1 -n)(2p+1 +n)}{4p(p+1)}\mu^\frac{n-1}{4} \nn
\eea
for $n$ odd and zero when $n$ is even. Using standard $\Gamma$-function identities it is then easy to show that the above equation is consistent with our computation \eqref{1pointfromloop}.

\section{Liouville cylinder amplitude}
\label{Cylindercalc}
The full cylinder amplitudes were given in a usable form recently by \cite{Gesser:2010fi}. We have obtained similar results independently and reproduce  them here for the purpose of completeness. The cylinder amplitude between a $(k,l)$ and $(r,s)$ brane, $\C{Z}(k,l;\sigma_1|r,s;\sigma_2)$, is computed by the integral
\bea \C{Z}(k,l;\sigma_1|r,s;\sigma_2) = \int^{\infty}_0 d\tau\C{Z}_{\mathrm{ghost}}(\tau) \C{Z}_{\mathrm{liouville}}(\tau) \C{Z}_{\mathrm{matter}}(\tau), \eea
where $\tau$ is the modular parameter of the cylinder in the closed string channel. The partition function for the ghosts is well known
\bea \label{Cylinder1} \C{Z}_{\mathrm{ghost}}(\tau) =\eta(\B{q})^2,\eea
where $\B{q} = \exp(-2 \pi\tau)$ and $\eta(\B{q})$ denotes the Dedekind $\eta$-function. The Liouville and matter contributions are given by
\bea 
\C{Z}_{\mathrm{matter}} = \sum_{(a,b) \in E_{q,p}} \frac{S_{(k,l),(a,b)}S_{(r,s),(a,b)}}{S_{(1,1),(a,b)}} \chi_{(a,b)}(\B{q}),\\
\C{Z}_{\mathrm{liouville}} = \int^{\infty}_0 dP \Psi^{\dagger}_{\sigma_1}(P)\Psi_{\sigma_2}(P) \chi_P (\B{q}), 
\eea
where $E_{q,p}$ is the Kac table for the minimal model, $S$ is the relevant modular $S$-matrix, and $\chi_{(a,b)}(\B{q})$ and $ \chi_{P}(\B{q})$ are the characters for the minimal model primary field with Kac index $(a,b)$ and the non-degenerate Liouville primary field respectively. The characters are given by,
\bea
\chi_{(a,b)}(\B{q})&&=\frac{1}{\eta(\B{q})} \sum^{\infty}_{n=-\infty}\left( \B{q}^{(2pqn+qa-pb)^2/4pq}-\B{q}^{(2pqn+qa+pb)^2/4pq}\right),\\
\chi_P (\B{q}) &&= \frac{\B{q}^{P^2}}{\eta(\B{q})}.
\eea
We first concentrate on the $\C{Z}_{\mathrm{matter}}$. Recalling that the modular S-matrix for the $(p,q)$ minimal model is,
\bea S_{(k,l),(r,s)} &&= 2\sqrt{\frac{2}{pq}}(-1)^{1+lr+ks}\sin(\pi\frac{qkr}{p})\sin(\pi\frac{pls}{q})
\eea
the explicit expression for $\C{Z}_{\mathrm{matter}}$ is,
\bea
\C{Z}_{\mathrm{matter}}&&=-2\sqrt{\frac{2}{pq}} \sum^{p-1}_{a=1} \sum^{q-1}_{b=1} \Big[(-1)^{a(l+r+1)+b(k+s+1)}\frac{\sin(\pi\frac{qka}{p})\sin(\pi\frac{plb}{q})\sin(\pi\frac{qra}{p})\sin(\pi\frac{psb}{q})}{\sin(\pi\frac{qa}{p})\sin(\pi\frac{pb}{q})}\nn \\
&&\frac{1}{\eta(\B{q})} \sum^{\infty}_{n=-\infty}\left( \B{q}^{(2pqn+qa-pb)^2/4pq}-\B{q}^{(2pqn+qa+pb)^2/4pq}\right) \Big]
\eea
where we have used the symmetry of the Kac table to rewrite the limits on the $a$ and $b$ summation. Also, note that the quantity in the square brackets is symmetric in $a$ and $b$ and zero if $a=0$ or $b=0$. This allows us to rewrite it as,
\bea
\C{Z}_{\mathrm{matter}}&&=\frac{1}{\sqrt{2pq}} \sum^{p-1}_{a=1} \sum^{q-1}_{b=-(q-1)} \Big[\sin(\frac{\pi t}{p})\sin(\frac{\pi t}{q}) \times \\
&&U_{k-1}(\cos\frac{\pi t}{p})U_{r-1}(\cos\frac{\pi t}{p})U_{l-1}(\cos\frac{\pi t}{q})U_{s-1}(\cos\frac{\pi t}{q}) \frac{1}{\eta(\B{q})} \sum^{\infty}_{n=-\infty}\B{q}^{(2pqn+t)^2/4pq}\Big] \nn
\eea
where $t=qa+pb$. Substituting these expressions into \eqref{Cylinder1} gives,
\bea
&&\C{Z}= \frac{1}{\sqrt{2pq}}\left[ \int^{\infty}_0 dP \frac{4\pi^2 \cos(2\pi\sigma_1 P)\cos(2\pi\sigma_2 P)}{\sinh(2\pi b P)\sinh(\frac{2\pi P}{b})} \right] \nn\\
&& \sum^{p-1}_{a=1} \sum^{q-1}_{b=-(q-1)} \Big[\sin(\frac{\pi t}{p})\sin(\frac{\pi t}{q}) U_{k-1}(\cos\frac{\pi t}{p})U_{r-1}(\cos\frac{\pi t}{p})U_{l-1}(\cos\frac{\pi t}{q})U_{s-1}(\cos\frac{\pi t}{q})\nn\\
&&\int^{\infty}_{0}d\tau \sum^{\infty}_{n=-\infty}\B{q}^{(2pqn+t)^2/4pq+P^2}\Big]
\eea
Focusing our attention on the integral over the moduli, we compute,
\bea
\int^{\infty}_{0}d\tau \sum^{\infty}_{n=-\infty}\B{q}^{(2pqn+t)^2/4pq+P^2}&&=-\frac{1}{2\pi}\sum^{\infty}_{n=-\infty}\left[ \frac{1}{(2pqn+t)^2/4pq+P^2} \right] \nn \\
&&=\frac{1}{2\sqrt{pq} P} \frac{\sinh(\frac{2\pi P}{\sqrt{pq}})}{\cos{\frac{\pi t}{pq}}-\cosh{\frac{2\pi P}{\sqrt{pq}}}}
\eea
which gives,
\bea
\label{Zintegrala}
\C{Z}= \frac{2\pi^2 }{\sqrt{2}pq} \int^{\infty}_0 \frac{dP}{P} \frac{\cos(2\pi\sigma_1 P)\cos(2\pi\sigma_2 P)\sinh(\frac{2\pi P}{\sqrt{pq}})}{\sinh(2\pi b P)\sinh(\frac{2\pi P}{b})} F_{k,l,r,s}(\frac{2\pi i P}{\sqrt{pq}})
\eea
where $F_{k,l,r,s}(z)$ is given by,
\bea 
\label{Fklrs}
F_{k,l,r,s}(z)=\sum^{\lambda_p(k,r)}_{\eta=|k-r|+1,2} \sum^{\lambda_q(l,s)}_{\rho=|l-s|+1,2} \sum^{p-1}_{a=1} \sum^{q-1}_{b=-(q-1)} &&\Bigg[\sin(\frac{\pi t}{p})\sin(\frac{\pi t}{q}) \times \nn \\ 
&&U_{\eta-1}(\cos\frac{\pi t}{p})U_{\rho-1}(\cos\frac{\pi t}{q}) \nn \frac{1}{\cos{\frac{\pi t}{pq}}-\cos{z}}\Bigg]\\
\eea
where $\lambda_p(k,r)=\mathrm{min}(k+r-1,2p-1-k-r)$. To compute \eqref{Zintegrala} we first note that the integrand is symmetric in $P$ and so we can extend the integral to the whole real line. Without loss of generality we assume $\sigma_1 > \sigma_2$, and break up the factor of $\cos(2\pi\sigma_1 P)$ in \eqref{Zintegrala} into its exponentials. For the integral containing, $\exp(2\pi i \sigma_1 P)$ we can close the integral along the real line by a semi-circle to produce the contour $C_+$ in the upper-half plane, for the other term we use a contour, $C_-$ in the lower-half plane. However, the integral using the contour in the lower-half plane, by a change of variables from $P$ to $-P$ is equal to the integral using the contour in the upper half-plane, we therefore get,
 \bea
\label{Zintegralb}
\C{Z}(k,l;\sigma_1|r,s;\sigma_2)= A_C\oint_{C_+} dP\left[ G(P,\sigma_1,\sigma_2) F_{k,l,r,s}(\frac{2\pi i P}{\sqrt{pq}})\right]
\eea
The poles of $F_{k,l,r,s}(\frac{2 \pi i P}{\sqrt{pq}})$ occur at $P \in S_F = \{\frac{n i}{2\sqrt{p q}}:n\in \BB{Z}, n \neq 0 \mod p, n \neq 0 \mod q\}$. The residue of $F_{k,l,r,s}(\frac{2 \pi i P}{\sqrt{pq}})$ at $P \in S_F$ is,
\bea
\label{Fresidues}
\res(F_{k,l,r,s}(\frac{2 \pi i P}{\sqrt{pq}}); P = \frac{i n}{2\sqrt{p q}}) =&& -2\frac{\sqrt{p q}}{2 \pi i} \sin(\frac{\pi n}{p})\sin(\frac{\pi n}{q}) U_{k-1}(\cos\frac{\pi n}{p})U_{r-1}(\cos\frac{\pi n}{p}) \times \nn \\ 
&&U_{l-1}(\cos\frac{\pi n}{q})U_{s-1}(\cos\frac{\pi n}{q})\frac{1}{\sin \frac{\pi n}{pq}}
\eea
The residues of the $G(P)$ are,
\bea
\label{Gresidues}
&&2\pi i\res\Big[G(P);P=i\epsilon\Big] = (\frac{1}{2\epsilon}-\pi \sigma_1)\frac{1}{\sqrt{pq}} \\
&&2\pi i\res\Big[G(P);P=\frac{inp}{2\sqrt{pq}}\Big]=\frac{2}{n} (-1)^n e^{-\frac{\pi p n \sigma_1}{\sqrt{pq}}} T_n(x_2) \frac{\sin(\pi n /q)}{\sin(\pi p n /q)}\nn \\
&&2\pi i\res\Big[G(P);P=\frac{inq}{2\sqrt{pq}}\Big]=\frac{2}{n} (-1)^n e^{-\frac{\pi q n \sigma_1}{\sqrt{pq}}} T_n(\tilde{x}_2) \frac{\sin(\pi n /p)}{\sin(\pi q n /p)}\nn \\
&&2\pi i\res\Big[G(P);P=\frac{inpq}{2\sqrt{pq}}\Big]=\frac{2}{npq} (-1)^{n(p+q+1)} e^{-\frac{\pi p q n \sigma_1}{\sqrt{pq}}} T_{qn}(x_2)\nn 
\eea

Finally, a useful identity for computing the sums found in the computation of the cylinder amplitudes is,
\bea
\label{sumident}
\sum^{\infty}_{t=1}\frac{4}{t}\gamma^t e^{-\alpha t}&&\cosh\beta t\cos\phi t =\nn \\
&&2\alpha - \ln(2(\cosh^2\alpha+\cosh^2\beta-2\gamma \cosh\alpha\cosh\beta\cos\phi-\sin^2\phi))
\eea


\section{Loop equations}
\label{MMloops}
To compute the resolvent $W_X$ we make the following change of variable in \eqref{ZXY},
\bea
S = M_1+M_2+1+c, \qquad
X = \frac{(1 - \alpha)}{2}M_1-\frac{(1 + \alpha)}{2}M_2-(1 + c) \frac{\alpha}{2}
\eea
The partition function in these new variables takes the form,
\bea
\mathcal{Z}=\int [dS dX] \exp{\left[-\frac{N}{g}\Tr\left(X^2 S+\alpha XS^2+V(S)\right)\right]}
\eea
where,
\beq 
V(S)=\frac{1}{12}(1+3\alpha^2)S^3-\frac{c}{2}S^2+\quarter(3c-1)(1+c)S
\eeq
This explains why we chose the parameterisation in which $W_{M_1+M_2}$ is awkwardly at $\alpha = \infty$; we can compute it by computing $W_S$ instead. As one would expect this matrix model is equivalent to an $O(1)$ model, as demonstrated by making the change of variable, $X\rightarrow (Z-\alpha S)/2$ which results in,
\bea
\Z=\int [dS dZ] \exp{\left[-\frac{N}{g}\Tr\left(Z^2 S/4+U(S)\right)\right]}
\eea
where all dependence on $\alpha$ is absorbed into the new potential $U(S)$. This shows that any correlation function that is odd in $Z=2X+\alpha S$ is zero, which will be of use later.

The resolvents can be obtained by generating an appropriate set of loop equations. We begin by obtaining the resolvents $W_X$ and $W_S$. Consider the following change of variables in the matrix model,
\bea
X \rightarrow X+\epsilon\frac{1}{z-S} 
\eea
this gives the loop equation,
\bea
\label{LE1d}
W^X{}_S(z)=\frac{\alpha}{2} - \frac{\alpha z}{2} W_S(z)
\eea
Now consider,
\bea
\label{LE1a}
X &\rightarrow X+\epsilon \left( \frac{1}{x-X} \frac{1}{z-S}+\frac{1}{z-S} \frac{1}{x-X} \right)\\
\label{LE1b}
S &\rightarrow S+\epsilon \left( \frac{1}{x-X} \frac{1}{z-S}+\frac{1}{z-S} \frac{1}{x-X} \right)\\
\label{LE1c}
X &\rightarrow X+\epsilon \left( \frac{1}{z-S}\frac{1}{x-X}\frac{1}{-z-S}+h.c \right)
\eea
Transformation \eqref{LE1a} together with \eqref{LE1b} and \eqref{LE1d} gives,
\bea
\label{LE1}
-\frac{1}{g} \Omega^P_X(z,x)=\Omega_S(z,x) \Omega_{SX}(z,x)+\frac{1}{N^2}\left(W_{SX;S}(z,x;z)-\alpha W_{SX;X}(z,x;x) \right)
\eea 
where we have introduced the following functions,
\bea \nonumber
&&\Omega_S(z,x) = W_S(z)-\alpha W_X(x) - \frac{1}{g} (x^2-\alpha^2 z^2 + V'(z))\\ 
\label{omegadef}
&&\Omega_{SX} = W_{SX}(z,x)+\frac{1}{g}(x-\frac{\alpha z}{2})\\ \nonumber 
&&\Omega^P_X =W^P_X(z,x)+\frac{\alpha}{2}+\frac{1}{g}(x-\frac{\alpha z}{2})(x^2-\alpha^2 z^2 + V'(z))
\eea
where $P=Q-\alpha^2 S+\alpha x -\frac{3\alpha^2 z}{2}$ and $Q = V'(z)-V'(S)$.

Finally, the third transformation \eqref{LE1b} gives,
\bea
\label{LE2}
\Omega_{SX}(z,x) \Omega_{SX}(-z,x)=&&\frac{1}{g} \left(W_S(z)+W_S(-z)+\alpha W_X(x)+\frac{1}{g}(x^2-\frac{\alpha^2 z^2}{4}) \right)\\ \nn
-&&\frac{1}{N^2}W_{SX;SX}(z,x;-z,x)
\eea
Eliminating $\Omega_{SX}(z,x)$ between this and (\ref{LE1}) gives,
\bea
\nonumber
&&\frac{1}{g}\Omega^P_{X}(z,x) \Omega^P_{X}(-z,x)=\Omega_S(z,x)\Omega_S(-z,x) \left(W_S(z)+W_S(-z)+\alpha W_X(x)+\frac{1}{g}(x^2-\frac{\alpha^2 z^2}{4}) \right)\\ 
&&-\frac{1}{N^2} \Big(g \Omega_S(z,x)\Omega_S(-z,x)W_{SX;SX}(z,x;-z,x) + \Omega^P_{X}(z,x) W_{SX;S}(-z,x;-z) \label{LEwithN} \\ \nonumber
&&-\alpha \Omega^P_{X}(z,x) W_{SX;X}(-z,x;x) +\Omega^P_{X}(-z,x)\Big(W_{SX;S}(z,x;z)-\alpha W_{SX;X}(z,x;x)\Big) \Big)
\eea
By substituting in the large $N$ expansion for all resolvents appearing in this expression we can obtain a recursive definition for the genus $h$ resolvents. The genus zero resolvents satisfy:
\bea
\label{mastereqn}
\frac{1}{g}{\Omega^P_{X}}^{(0)}(z,x) {\Omega^P_{X}}^{(0)}(-z,x)=&& \\ \nn
{\Omega_S(z,x)}^{(0)}{\Omega_S(-z,x)}^{(0)}&&\left({W_S}^{(0)}(z)+{W_S}^{(0)}(-z)+\alpha {W_X}^{(0)}(x)+\frac{1}{g}(x^2-\frac{\alpha^2 z^2}{4}) \right)
\eea

This is an important equation and all subsequent loop equations will be related to this through derivative-like operations such as the loop insertion operator. 
It is important to note that the LHS is polynomial in $z$ and so if we expand the RHS in $x$ about any point, the resulting Laurent expansion must have coefficients that are equal to a polynomial function of $z$. A convenient point to choose for this expansion is infinity as the definition of the resolvent then  coincides with the Laurent expansion, whose coefficients are the yet to be determined quantities, $\avg{ \Tr{X^n}}$. Thus we produce a number of equations containing ${W_S}^{(0)}(z)$, 
\bea
\label{eqngen}
\oint_{C_{\infty}} \frac{dx}{2\pi i x^{n+1}} {\Omega_S(z,x)}^{(0)}&&{\Omega_S(-z,x)}^{(0)} \times \\ \nn &&\left({W_S}^{(0)}(z)+{W_S}^{(0)}(-z)+\alpha {W_X}^{(0)}(x)+\frac{1}{g}(x^2-\frac{\alpha^2 z^2}{4})\right) = p_n(z)
\eea
where $C_{\infty}$ is a contour around infinity, $p_n(z)$ is a polynomial in $z$ with a finite number of unknown coefficients, $n$ is an integer and the integral merely picks out the $n$th coefficient of the expansion. From the large $x$ behaviour of $W_X(x) \sim \frac{1}{x}$ it is easy to show that for $n>6$ the LHS is zero and for $n\geq 4$ the LHS yields an expression with no dependence on ${W_S}^{(0)}$. The non-trivial cases occur for $n=0$ and $n=2$ which generate enough equations to solve for ${W_S}^{(0)}(z)$. The large $z$ expansion of equations obtained for $n<0$ give relations among the quantities $\avg{\Tr{S^n}}$. The explicit equation obtained for ${W_S}^{(0)}(z)$ is,
\bea
\label{WSeqn}
{W_S}^{(0)}(z)^3&+& \frac{1}{g}\Big(\frac{3 z^2 \alpha^2}{4} + V'(-z) - 2 V'(z)\Big) {W_S}^{(0)}(z)^2\\
&+& \frac{1}{4 g^2}\Big(V'(-z) \Big(3 z^2 \alpha^2 - 4V'(z)\Big)-3 z^2 \alpha^2 V'(z) + 4 V'(z)^2 + P_2(z)\Big){W_S}^{(0)}(z) \nn\\
&+&\frac{1}{4 g^3}(P_0(z) + (z^2 \alpha^2 - V'(z)) P_2(z)) = 0\nn
\eea
where,
\bea
\label{Pdef}
P_i(z)=-p_i(z)+(\textrm{Coefficient of }({W_S}^{(0)}(z))^0\textrm{ in \ref{eqngen} for } n=i)
\eea
The unknown constants in $p_n(z)$ are not all independent and they may be found in terms of the constants $\langle \Tr{S^n} \rangle$ by expanding the RHS of \eqref{eqngen} about $z=\infty$ and equating $p_n$ to the polynomial part of the Laurent expansion. The requirement that the singular part of the expansion vanishes again gives relations between the quantities $\avg{\Tr{S^n}}$. 

To find ${W_X}^{(0)}$, we again consider \eqref{mastereqn}, however we now know ${W_S}^{(0)}$ in terms of a finite number of unknown constants. On the LHS we have the function ${\Omega^P_{X}}^{(0)}(z,x)$ where $P$ is a polynomial in $z$ and $S$ as defined in \eqref{omegadef}, whose highest power for both $z$ and $S$ is $d-1$ and so it contains $d$ unknown function of the form $W^{S^n}_X(x)$ for $0 \leq n <d$.

We can generate a system of equations for these resolvents by again expanding both sides of the loop equation about infinity but this time in terms of $z$. This results in,
\bea
\label{WXgen}
&&\oint_{C_{\infty}} \frac{dz}{z^{n+1}}\bigg[\frac{1}{g}{\Omega^P_{X}}^{(0)}(z,x) {\Omega^P_{X}}^{(0)}(-z,x)-\\ \nonumber
&&{\Omega_S(z,x)}^{(0)}{\Omega_S(-z,x)}^{(0)} \left({W_S}^{(0)}(z)+{W_S}^{(0)}(-z)+\alpha {W_X}^{(0)}(x)+\frac{1}{g}(x^2-\frac{\alpha^2 z^2}{4}) \right)\bigg]=0
\eea

Again we may vary $n$ to generate different equations, however we now generate non-trivial equations for all $0 \leq n <d$. Which is enough to solve for all unknown resolvent functions appearing in ${\Omega^P_{X}}^{(0)}(z,x)$. However, unlike ${{W_S}^{(0)}}$ for which we obtained an equation for general $V(x)$, we were unable to do this for the equations for $W^{S^n}_X(x)$. We also note that for the equation giving ${W_X}^{(0)}$, the highest power of ${W_X}^{(0)}$ appearing depends on the order of $V(x)$.

The approach thus outlined gives an algorithmic solution to this matrix model with an arbitrary potential and results in algebraic equations for all resolvents.

We now apply the above procedure to the case when the potential is \eqref{VSX}. There are two unknown constants in $P_{0}(z)$ and $P_{2}(z)$ corresponding to $\langle \Tr{S} \rangle$ and $\langle \Tr{S^2} \rangle$. Their values are determined by the requirement that \eqref{WSeqn} is a genus zero curve - which, in this case, is equivalent to the one-cut assumption. 

For ${W_{X}}^{(0)}$ we can generate two equations from \eqref{WXgen}, for $n=0$ and $n=1$, containing ${W_{X}}^{(0)}$ and ${W^S_{X}}^{(0)}$. The equation for ${W_{X}}^{(0)}$ resulting from the elimination of ${W^S_{X}}^{(0)}$ has a number of interesting properties. Generically it is quartic in ${W_{X}}^{(0)}$, however it reduces to a cubic for $\alpha = \pm 1$ and it reduces to a quadratic of $({W_{X}}^{(0)}+ \textrm{polynomial in } x)^2$ for $\alpha=0$. Because of this the scaling limit for generic $\alpha$ does not hold for these particular values of $\alpha$ and must be taken separately.

As was discussed before, the resolvents corresponding to the spin $+$ and $-$ Cardy states are ${W_{M_1}}^{(0)}$ and ${W_{M_2}}^{(0)}$ respectively, which may be computed directly from \eqref{ZXY} and are clearly identical. However a faster method, since we already have ${W_{X}}^{(0)}$, is to set $\alpha=-1$ implying $X=M_1+\half(1+c)$. 

It is easy to generate loop equations for a general 2-loop amplitudes of the form $W_{S;H}$ and $W_{X;H}$ where $H$ is a string of $S$ and $X$ matrices of length $\mathcal{N}_H$. We can write $H$ as, $H=\prod^{\mathcal{N}_H}_{n=1}\chi_n$, where $\chi_n$ is a matrix defined by,
\begin{equation}
\chi_n = \left\{
\begin{array}{c l}
X, & n\in\C{I}^H_X \\
S, & n\in\C{I}^H_S
\end{array}
\right .
\end{equation}
and $\mathcal{I}^H_X$ and $\mathcal{I}^H_S$ are disjoint indexing sets. For such a product, the notation $H_{(i,j)}$ is defined as ,
\beq
H=\prod^j_{n=i}\chi_i
\eeq
\noindent The loop equations for $W_{S;H}$, can be generated by the following changes of variables.
\bea
\label{LE4d}
X &&\rightarrow X+\epsilon\frac{1}{z-S} \Tr{H(S,X)} \\
\label{LE4a}
X &&\rightarrow X+\epsilon \left( \frac{1}{z-S}\frac{1}{x-X}+\frac{1}{x-X}\frac{1}{z-S} \right) \Tr{H(S,X)} \\
\label{LE4b}
S &&\rightarrow S+\epsilon \left( \frac{1}{z-S} \frac{1}{x-X}+\frac{1}{x-X} \frac{1}{z-S} \right) \Tr{H(S,X)} \\
\label{LE4c}
X &&\rightarrow X+\epsilon \left( \frac{1}{z_1-S}\frac{1}{x-X}\frac{1}{z_2-S}+h.c \right) \Tr{H(S,X)}
\eea
The loop equations generated by $X\rightarrow X+\epsilon F(X,S)\Tr{H}$, may by obtained from the loop equations generated by $X\rightarrow X+\epsilon F(X,S)$ by the following procedure,
\bea
\label{looprules}
W_{{G_1}(X,S)} &&\rightarrow W_{{G_1}(X,S);H} \nn \\ 
W_{{G_1}(X,S)} W_{{G_2}(X,S)} &&\rightarrow W_{{G_1}(X,S)} W_{{G_2}(X,S);H}+W_{{G_2}(X,S)} W_{{{G_1}(X,S)};H} \nn \\ 
\frac{1}{N^2}(...)&& \rightarrow \frac{1}{N^2}(...) + \sum_{i \in \C{I}^H_X} W_{H_{(1,i)}F(X,S) H_{(i,\C{N}_H)}}(z)
\eea
where ${G_i}(X,S)$ is any function of $X$ and $S$. An equivalent set of rules applies for loops equations generated by $S\rightarrow S+\epsilon F(X,S)\Tr{H}$. The resulting loop equations can then be solved by using the same method as used to solve the one-loop amplitude equations. For example, the two loop amplitude $W_{S;S}(z;u)$ may be obtained by choosing $H=(u-S)^{-1}$ in \eqref{LE4d}, \eqref{LE4a}, \eqref{LE4b} and \eqref{LE4c}. Applying the rules in \eqref{looprules} we get,
\bea
-\frac{1}{g}\Omega^P_{X;S}(z,x;u)&=&\Omega_{SX;S}(z,x;u)\Omega_{S}(z,x)+ \nn \\ 
&&\Omega_{SX}(z,x)\Omega_{S;S}(z,x;u)+\partial_u \left[\frac{W_{SX}(u)-W_{SX}(z)}{u-z}\right]
\eea
and
\bea
\Omega_{SX}(z,x)\Omega_{SX;S}(-z,x;u)&+&\Omega_{SX;S}(z,x,u)\Omega_{SX}(-z,x) = \nn \\
&&\frac{1}{g} \left(W_{S;S}(z;u)+W_{S;S}(-z;u)+\alpha W_{X;S}(x;u)\right)
\eea
By eliminating $\Omega_{SX;S}(z,x,u)$ we obtain an equation relating $W_{S;S}(z;u)$, $W_{S;S}(-z;u)$ and $W_{X;S}(x;u)$. If we arrange the resulting equation so that the LHS is polynomial in $z$, we may expand the RHS in $x$ about $x=\infty$ to generate a number of equations for $W_{S;S}(z;u)$ and $W_{S;S}(-z;u)$. This is exactly the same procedure we used to compute the disc amplitudes in which the resulting expression contained a number of unknown constants, which for the equal potential case were $\avg{\Tr{S}}$ and $\avg{\Tr{S^2}}$. However here we get an expression for $W_{S;S}(z;u)$ in terms of unknown functions of the form $W^{S}{}_{;S}(u)$ and $W^{S^2}{}_{;S}(u)$. These unknown functions may be found by computing the expansion of $W_{S;S}$ at $z=\infty$ and, by the symmetry of $W_{S;S}$, equating it to the expansion at $u=\infty$. The expressions for $W^{S}{}_{;S}(z)$ and $W^{S^2}{}_{;S}(z)$ then depend on a number of unknown constants of the form, $\avg{\Tr{S^m}\Tr{S^n}}_c$, not all of which are independent, as can be seen by considering the large $z$ expansion of the $W^{S}{}_{;S}(z)$ and $W^{S^2}{}_{;S}(z)$. In the end the independent unknown quantities correspond to, $\avg{\Tr{S}\Tr{S}}_c$ and $\avg{\Tr{S}\Tr{S^2}}_c$ which can be fixed by requiring that $W^{S}{}_{;S}(z)$ has no singularities besides those at the branch points of $W_S$. The loop equations then give $W_{X;S}$ in terms of $W_{S;S}$. Given that the continuum limit for generic values of $\alpha$ does not give any more information we choose to set to $\alpha = -1$ without loss of generality. Computing the scaling form of $W_{S;X}$ gives \eqref{WSXeqn} and the other two loop amplitudes are found in a similar manner

In order to compute the $1/N$ corrections it is clear from \eqref{LEwithN} that $W_{SX;S}$, $W_{SX;X}$ and $W_{SX;SX}$ need to be calculated. The method for calculating $W_{SX;S}$ was given in the preceding section. The calculation of $W_{SX;X}$ and $W_{SX;SX}$, follows a similar procedure however the resulting loop equations contain new amplitudes of the form $W_{SX..SX}$. The calculation of these new quantities does not pose a great difficulty with the appropriate change of variables giving loop equations very similar to the ones presented previously. 

Once these quantities have been calculated and substituted in \eqref{LEwithN}, we may again expand both sides around $x=\infty$ and then $z=\infty$ to find an expression for ${W_S(z)}^{(1)}$ and ${W_X(x)}^{(1)}$. Again this procedure introduces a new unknown constant corresponding to the $\frac{1}{N}$ correction to $\avg{\Tr{S}}$ which may be determined by requiring that ${W_S(z)}^{(1)}$ does not possess any poles besides those at the branch points of ${W_S(z)}^{(0)}$. 
The result of the calculation is \eqref{DWHresolv}

\end{document}